\documentclass[aps,prc,twocolumn,floatfix,superscriptaddress,showpacs]{revtex4-1}
\usepackage{graphicx}
\usepackage[pdftex,colorlinks,citecolor=blue,bookmarks]{hyperref}
\usepackage{bm}
\usepackage{longtable}
\begin{document}
\title{Toward global beyond-mean-field calculations of nuclear masses and low-energy spectra}

\author{Tom\'as R. Rodr\'iguez}
\affiliation{Departamento de F\'isica Te\'orica, Universidad
  Aut\'onoma de Madrid, E-28049 Madrid, Spain}
\affiliation{Institut f\"ur Kernphysik, Technische Universit\"at
  Darmstadt, D-64289 Darmstadt, Germany}  
\author{Alexander Arzhanov} 
\affiliation{Institut f\"ur Kernphysik, Technische Universit\"at
  Darmstadt, D-64289 Darmstadt, Germany}  
\author{Gabriel Mart\'inez-Pinedo}
\affiliation{Institut f\"ur Kernphysik, Technische Universit\"at
  Darmstadt, D-64289 Darmstadt, Germany}  
\affiliation{GSI Helmholtzzentrum f\"ur Schwerionenforschung,
  Planckstra{\ss}e~1, 64291 Darmstadt, Germany}
\date{\today} 

\begin{abstract}
Self-consistent mean field (MF) and beyond-mean-field (BMF)
calculations of masses, separation energies and $2^{+}_{1}$ excitation
energies of even-even nuclei where experimental data is available are
presented. The functionals used are based on the Gogny D1S and D1M
parametrizations and the method includes beyond-mean-field corrections
coming from both axial quadrupole shape mixing and symmetry restorations
without assuming gaussian overlap approximations. A comparison between
mean field, beyond-mean-field approaches and the experimental data is
provided. Additionally, the convergence of the results and the possible reduction of the magic shell gaps by including BMF effects are also discussed.
\end{abstract}
\maketitle

\section{Introduction}
\label{Intro}

Nuclear binding energies are one of the most relevant quantities that
define the atomic nucleus. Apart from their intrinsic interest, the
nuclear masses are specially important in determining the
nucleosynthesis processes that occur in astrophysical
environments. For example, they determine the limits of existence of
the nucleus as a bound system of protons and neutrons (drip lines),
the $Q$-values for $\beta$ and $\alpha$ decays, the particle
separation energies that are needed to compute capture/emission rates,
etc. Therefore, a great effort in measuring masses of very exotic
(short-lived) nuclei with high precision is currently made with the
use of trapping and/or storage rings techniques~\cite{PR_425_1_2006}.  

Despite this significant progress, and the one expected in the near
future with the arrival of new facilities worldwide, many nuclei will
not be experimentally reachable. Particularly important are those
belonging to neutron rich regions that are relevant to better
constraint the rapid neutron capture (r-process)
nucleosynthesis, the mechanism behind the origin of more than the half
of the elements beyond iron in the universe. Hence, nuclear models
able to compute with high precision the known masses, as well as to
provide reliable extrapolations, are very much demanded.

Current theoretical nuclear mass tables are provided mainly by three
different approaches: (a) Microscopic-macroscopic (mic-mac) methods
which improve the original liquid drop formula with microscopic
corrections. The most commonly used of this kind are the finite range
droplet model (FRDM)~\cite{ADNDT_59_185_1995,PRL_108_052501_2012} or,
more recently, the Weizs\"acker-Skyrme (WS) mass
model~\cite{PRC_81_044322_2010,PLB_734_215_2014}. (b) Duflo-Zuker (DZ)
approach based a functional of occupation numbers guided by the
interacting shell model method~\cite{PRC_52_R23_1995}. (c) Microscopic
methods based on Hartree-Fock-Bogoliubov (HFB) approaches with Skyrme
(see Ref.~\cite{PRC_88_024308_2013,PRC_88_061302_2013} and references
therein) and Gogny functionals~\cite{PRL_102_242501_2009}. The above
methods have reached a root mean square (RMS) deviation from
data~\cite{AW2003,AW} of 0.57 MeV (FRDM), 0.298 MeV (WS), 0.36 MeV
(DZ), 0.51 MeV (HFB-Skyrme) and 0.798 MeV (HFB-Gogny).

This impressive agreement to the global behavior of the known masses
is not sufficient to have a full confidence in the extrapolations to
unknown regions since large deviations in the predictions of the
different models are found. In addition, even if the overall
performance is similar, some local deviations between the models in
sensitive regions can largely affect the results of nucleosynthesis
simulations~\cite{PRC_83_045809_2011,EPJA_48_184_2012}. Therefore,
theoretical models should be improved to reduce such uncertainties.  
In particular, \textit{ab-initio} calculations are becoming available
for medium mass systems although currently neither the range of
applicability nor the accuracy reached by such methods are
good enough to be applied to astrophysical
purposes~\cite{PRL_107_072501_2011,PRC_69_054320_2004,PRL_110_242501_2013,PRC_87_011303_2013,PRL_110_022502_2013,Nature_502_207_2013,PRL_105_032501_2010}. Hence,
in the short-mid term energy density functionals are still
the most promising microscopic methods to compute nuclear properties
in the whole nuclear chart with the required accuracy.  
These methods have been improved significantly in the last years by
taking into account not only global fits to all known masses but also
some constraints coming from \textit{ab-initio} calculations in 
infinite nuclear
matter~\cite{PRC_88_024308_2013,NPA_812_72_2008,PLB_668_420_2008}. 

However, a mean field (HFB) approach is commonly used to solve the
nuclear many-body problem and beyond-mean-field (BMF) corrections have
been included phenomenologically to mimic rotational and vibrational
corrections~\cite{PRC_75_064312_2007}. Nevertheless, in the last
decade, a better treatment of such BMF correlations of even-even
nuclei using symmetry restorations and configuration mixing methods
have been implemented with Skyrme, Relativistic and Gogny
interactions~\cite{RMP_75_121_2003}. These improvements have allowed
the study of the appearance/disappearance of shell closures or shape
mixing/coexistence/transitions phenomena, for example. Concerning
nuclear masses, these BMF correlations have been computed globally
using particle number and angular momentum projection and generator
coordinate method (GCM) with Skyrme
interactions~\cite{PRC_73_034322_2006,PRC_78_054312_2008} and using
the five dimensional collective hamiltonian (5DCH) with Gogny
~\cite{PRL_102_242501_2009,PRC_81_014303_2010} and, more recently, Relativistic~\cite{PRC_91_027304_2015} interactions.  Due to
the large computational burden of performing GCM with
particle number and angular momentum restorations, these pioneer
global surveys were carried out assuming gaussian overlap approximations (GOA).

In this paper we present the results of global calculations for
even-even nuclei performed with the Gogny
D1S~\cite{NPA_428_23_1984,CPC63_365_1991} and
D1M~\cite{PRL_102_242501_2009} interactions using 
implementations of the GCM and quantum number projection methods without gaussian overlap
approximations. Additionally, we discuss the convergence of the results -both HFB and BMF- as a function of the number of oscillator shells included in the basis where the many-body wave functions are expanded.

Results for odd systems including such advanced many-body techniques
are still not available with the Gogny interaction although some
preliminary calculations in a single nucleus have been recently
reported with Skyrme forces~\cite{PRL_113_162501_2014}. Additionally,
in this survey we assume axial and parity symmetry conserving
intrinsic states, exploring explicitly the axial quadrupole degree of
freedom by performing one-dimensional GCM calculations.  Including
triaxiality and/or octupolarity within the present framework increases
prohibitively the computational time and it is beyond the scope of the
present study. Nevertheless, global calculations including the
  triaxial~\cite{PRL_97_162502_2006,PRL_102_242501_2009,PRC_81_014303_2010},
  and octupole~\cite{PRC_84_054302_2011,arXiv:1408.6941} degrees of
  freedom within less involved many-body methods have been
  reported. In particular, 159 out of 5900 nuclei with triaxial ground
  states are found in the FRDM model~\cite{PRL_97_162502_2006}. The
  largest differences in binding energies, $\lesssim 0.6$ MeV, are
  obtained in regions around $^{108}$Ru and $^{140}$Gd. In
  Ref.~\cite{PRC_81_014303_2010} most of the nuclei are predicted to
  be axial symmetric (spherical, prolate and oblate) at the HFB level
  with Gogny D1S. BMF correlations within the 5DCH bring all the
  nuclei toward triaxial shapes --in average-- even though in the HFB approximation 
  are predicted spherical and/or axial symmetric.
  On the other hand, following
  Ref.~\cite{arXiv:1408.6941}, only a few out of 818 nuclei calculated
  with Gogny D1S-D1N-D1M between $8\leq Z\leq110$ are octupole
  deformed in the HFB ground state. These nuclei are obtained around
  Ra, Ba and Zr region and the energy gain with respect to the
  octupole symmetric states are~$\lesssim 1.2$~MeV. Further BMF
  correlations including parity projection and GCM along the
  $\beta_{3}$ degree of freedom give extra binding energies as large as $2.5$
  MeV in the regions showing octupole deformation at the HFB
  level. However, in the rest of nuclei these correlation energies are
  of the order of 1 MeV and vary smoothly with $N$ and $Z$, producing
  an almost constant shift in the total energies.

This document is organized as follows: First, a description of the
method used to compute masses is given in Sec.~\ref{Theo}. Then, the
results are discussed in Sec.~\ref{Res}. Finally, a brief summary and
outlook are presented in Sec.~\ref{Summ}. 

\section{Theoretical framework}
\label{Theo}

The total energy (negative in our convention) of a given nucleus is
calculated in the present work as the sum of two terms:
\begin{equation}
E=E_{\mathrm{HFB}}+\Delta E_{\mathrm{BMF}}
\label{benergy}
\end{equation}
where $E_{\mathrm{HFB}}$ is the Hartree-Fock-Bogoliubov (HFB) energy
(mean field) and $\Delta E_{\mathrm{BMF}}$ is a beyond-mean-field
(BMF) correction which includes particle number restoration, angular
momentum projection and axial quadrupole shape mixing within the
generator coordinate method (GCM)~\cite{RING_SCHUCK}. Both terms are
computed with the same underlying interaction, namely, Gogny D1S or
D1M parametrization.

For the sake of simplifying the notation, we express the theoretical
energy throughout this work as the expectation value of a hamiltonian
operator. However, it is important to point out that since the
effective interactions used here are density-dependent, this notation
is not truly rigorous and energy density functionals (EDFs) should be
defined instead of such expectation values. We refer to
Ref.~\cite{PRC_79_044318_2009} for a general discussion about this
topic and Refs.~\cite{NPA_709_201_2002,PRC_81_064323_2010} for the
particular choice of the corresponding EDFs used in this work.   

\subsection{Mean field (HFB) approach}\label{MF}

We start by reviewing briefly the HFB method~\cite{RING_SCHUCK}. In
this microscopic approach, based on the variational principle, the
many-body wave function of the atomic nucleus is searched among a set of trial wave functions that are defined as
quasiparticle vacua, $|\phi\rangle$: 

\begin{equation}
\hat{\beta}_{k}|\phi\rangle=0 \,\, \forall \,\, k
\label{HFBvar}
\end{equation}

Those quasiparticles are defined as the most general linear
combination of creation ($\hat{c}^{\dagger}_{l}$) and annihilation
($\hat{c}_{l}$) single particle operators: 
\begin{equation}
\hat{\beta}^{\dagger}_{k}=\sum_{l}U_{lk}\hat{c}^{\dagger}_{l}+V_{lk}\hat{c}_{l},
\label{HFBtrans}
\end{equation}
where $U$, and $V$ are the variational parameters. Since
Eq.~\ref{HFBtrans} breaks most of the symmetries of the original
interaction, in particular, the particle number symmetry, the HFB wave
function is constrained to have the correct mean value of the number
of particles. Therefore, the HFB equations~\cite{RING_SCHUCK} are
found by the condition: 
\begin{equation}
\delta\left(E^{'}_{\mathrm{HFB}}\left[|\phi\rangle\right]\right)_{|\phi\rangle=|\mathrm{HFB}\rangle}=0
\label{Ritz}
\end{equation}
with
\begin{equation}
E^{'}_{\mathrm{HFB}}\left[|\phi\rangle\right]=\langle\phi|\hat{H}|\phi\rangle-\lambda_{N}\langle\phi|\hat{N}|\phi\rangle-\lambda_{Z}\langle\phi|\hat{Z}|\phi\rangle,
\label{Eprime}
\end{equation}
where $\hat{H}$, $\hat{N}(\hat{Z})$ are the hamiltonian and the
neutron (proton) number operators, respectively; $|\phi\rangle=|\mathrm{HFB}\rangle$ is the
HFB solution obtained by solving the corresponding HFB equations (Eqs.~\ref{Ritz}-\ref{Eprime}); and $\lambda_{N(Z)}$ is a Lagrange multiplier which
ensures $\langle\phi|\hat{N}(\hat{Z})|\phi\rangle=N(Z)$. The
normalization $\langle\phi|\phi\rangle=1$ is also assumed.

In practical applications, the spherical harmonic oscillator (s.h.o.)
basis is usually chosen as the working basis where the quasiparticles
defined in Eq.~\ref{HFBtrans} are expanded~\cite{footnote}. The sum in such an
equation runs over an infinite number of s.h.o. states but this sum
must be truncated in actual implementations in computer codes. The
results should not depend on the choice of the basis and the
convergence of the results are obtained if a sufficiently large number
of major s.h.o. shells ($N_{s.h.o.}$) are included. However, the
computational burden increases significantly with the number of
oscillator shells and a compromise between a better convergence and a
reasonable computational time has to be considered (see discussion below). In the present
work, $E_{\mathrm{HFB}}=\langle\mathrm{HFB}|\hat{H}|\mathrm{HFB}\rangle$ in Eq.~\ref{benergy} is computed with
$N_{s.h.o.}=19$. 
\subsection{Beyond-mean-field (BMF) approach}

The second term in the energy ($\Delta E_{\mathrm{BMF}}$ in Eq.~\ref{benergy}) corresponds to corrections beyond the
mean field (HFB) approximation. In principle, the energy should be
computed by using BMF methods from the first place. However, these
methods are much more time consuming than the corresponding HFB and
the size of the s.h.o. basis used in this case is generally
smaller. In the present work, $N_{s.h.o.}=11$ has been chosen
for computing BMF effects. Hence, BMF correction is defined as: 
\begin{equation}
\Delta E_{\mathrm{BMF}}=E_{\mathrm{BMF}}(N_{s.h.o.}=11)-E_{\mathrm{HFB}}(N_{s.h.o.}=11)
\label{bmf_corr}
\end{equation}
This energy difference is less dependent on the number of oscillator shells included in the basis than the total energies separately, $E_{\mathrm{HFB}}$ and $E_{\mathrm{BMF}}$. We will study explicitly this point below.

In the above equation, $E_{\mathrm{BMF}}(N_{s.h.o.}=11)$ is computed
within a symmetry conserving configuration mixing (SCCM)
framework. The method contains simultaneous particle number and
angular momentum projection (PNAMP) of different intrinsic
Hartree-Fock-Bogoliubov-type states and a subsequent mixing
of these states performed within the generator coordinate method (GCM)
framework. In general, different shapes (axial and non-axial) and
collective coordinates can be included in the GCM
calculation. However, the addition of more generating coordinates
largely increases the computational time, especially if a triaxial
angular momentum projection is performed. Therefore, we have focused
on the mixing of axial deformed and parity conserving states. Further
extensions will be explored in a future study. Nevertheless,
  contrary to other BMF approaches like the 5DCH, this method is
  variationally consistent with the underlying HFB functional, i.e., lower
  total energies are always obtained when such correlations are taken
  into account. Furthermore, the more collective coordinates and
  symmetry restorations are included in the GCM the lower is the
  ground state energy obtained until the exact solution is eventually
  obtained. In this sense, the method is extensible to other degrees
  of freedom (triaxiality, octupolarity, pairing content, ...) without
  loosing its variational character. Finally, although the present
  calculations are restricted to axial, parity and time-reversal
  symmetric intrinsic states, the method can be applied to the whole
  nuclear chart. If a specific nucleus turns out to be, for instance,
  triaxial and/or octupole deformed, the amount of correlations
  obtained here will be smaller, though neither negligible nor
  meaningless, than the ones eventually obtained with including those
  degrees of freedom.
 
In the following, the particular
realization of the SCCM method used here is described step-by-step. 

\subsubsection{Variation after particle number projection method
  (PN-VAP)} 

In contrast to the previous section (Sec.~\ref{MF}), the particle
number projected energy is minimized instead of the HFB
one. This is the so-called variation after particle number projection (PN-VAP)~\cite{NPA_696_467_2001,RING_SCHUCK}. 
Furthermore, the set of intrinsic, HFB-type, trial wave functions, $|\phi(\beta_{2})\rangle$, are also constrained to have a given value of the axial quadrupole deformation, $\beta_{2}$. Hence, the variational
equation now reads as: 
\begin{equation}
\delta\left(E^{'}_{\mathrm{PN-VAP}}\left[|\phi\rangle\right]\right)_{|\phi\rangle=|\mathrm{PNVAP}\rangle}=0,
\label{RitzPNVAP}
\end{equation}
with $|\phi(\beta_{2})\rangle=|\mathrm{PNVAP}(\beta_{2})\rangle$ is the intrinsic wave function that minimizes the functional:
\begin{eqnarray}
E^{'}_{\mathrm{PN-VAP}}\left[|\phi\rangle\right]=\frac{\langle\phi|\hat{H}P^{N}P^{Z}|\phi\rangle}{\langle\phi|P^{N}P^{Z}|\phi\rangle}&-&\lambda_{q_{20}}\langle\phi|\hat{Q}_{20}|\phi\rangle.
\end{eqnarray}
Here, $P^{N(Z)}$ is the projector onto good number of neutrons
(protons)~\cite{RING_SCHUCK},$\hat{Q}_{20}=r^{2}Y_{20}(\theta,\varphi)$
is the axial quadrupole operator and the Lagrange multiplier
$\lambda_{q_{20}}$ ensures the condition:
$\lambda_{q_{20}}\rightarrow\langle\phi|\hat{Q}_{20}|\phi\rangle=q_{20}$. 
The quadrupole deformation parameter $\beta_{2}$ is related to
$q_{20}$ by: 
\begin{equation}
q_{20}=\pm\frac{\beta_{2}}{C};\,\,\,\,C=\sqrt{\frac{5}{4\pi}}\frac{4\pi}{3r_{0}^{2}A^{5/3}}
\end{equation}
being $r_{0}=1.2$ fm, $A$ the mass number and the plus (minus) sign
indicates prolate (oblate) shapes. 
Hence, the collective intrinsic deformation is well established within
this framework and this fact allows the description of the states in
the laboratory frame in terms of their intrinsic shapes unambiguously.

\subsubsection{Symmetry conserving configuration mixing (SCCM) method}

Once the set of intrinsic wave functions -$|\mathrm{PNVAP}(\beta_{2})\rangle$-
is obtained, the final states are built through the Ansatz provided by
the generator coordinate method (GCM)~\cite{RING_SCHUCK}. In this
framework, the states are assumed to be linear combinations of
particle number and angular momentum projected PN-VAP states: 
\begin{equation}
|\Psi^{I\sigma}\rangle=\sum_{\beta_{2}}g^{I\sigma}_{\beta_{2}}P^{I}_{00}P^{N}P^{Z}|\mathrm{PNVAP}(\beta_{2})\rangle
\label{GCM_state}
\end{equation}
where $I$ is the total angular momentum, $P^{I}_{KK'}$ the angular
momentum projector applied to axial symmetric intrinsic states
($K=K'=0$)~\cite{RING_SCHUCK} and $\sigma$ labels different states
obtained for a given value of $I$. The parameters
$g^{I\sigma}_{\beta_{2}}$ 
are determined by the Ritz variational principle which leads to the
Hill-Wheeler-Griffin (HWG) equation: 
\begin{widetext}
\begin{equation}
\delta \left(E^{I\sigma}_{\mathrm{SCCM}} \left[g^{I\sigma}_{\beta_{2}}\right]\right)
= 0\Rightarrow 
\sum_{\beta'_{2}}
\left(\mathcal{H}^{I}_{\beta_{2},\beta'_{2}} -
  E^{I\sigma}_{\mathrm{SCCM}} \,
  \mathcal{N}^{I}_{\beta_{2},\beta'_{2}}\right)g^{I\sigma}_{\beta'_{2}}
= 0  
\label{RitzGCM}
\end{equation}
\end{widetext}
The energy and norm overlap matrices are defined as:
\begin{eqnarray}
\mathcal{H}^{I}_{\beta_{2},\beta'_{2}}&=&\langle\mathrm{PNVAP(\beta_{2})}|\hat{H}P^{I}_{00}P^{N}P^{Z}|\mathrm{PNVAP}(\beta'_{2})\rangle\nonumber\\
\mathcal{N}^{I}_{\beta_{2},\beta'_{2}}&=&\langle\mathrm{PNVAP(\beta_{2})}|P^{I}_{00}P^{N}P^{Z}|\mathrm{PNVAP}(\beta'_{2})\rangle\nonumber\\
\end{eqnarray}
The resulting HWG equations -one for each value of the angular 
momentum- provide the energy levels $E^{I\sigma}_{\mathrm{SCCM}}$ and
collective wave functions defined in the $(\beta_{2})$ direction.  

Hence, the energy including symmetry restorations and shape mixing
within this framework is given by: 
\begin{equation}
E_{\mathrm{BMF}}(N_{s.h.o.}=11)=E^{I=0\sigma=1}_{\mathrm{SCCM}}
\end{equation}
Obviously, excited states, in particular
$E(2^{+}_{1})=E^{I=2\sigma=1}_{\mathrm{SCCM}}-E^{I=0\sigma=1}_{\mathrm{SCCM}}$
excitation energies, can be also calculated within the same framework.  

\begin{figure}[tb]
  \begin{center}
    \includegraphics[width=\columnwidth]{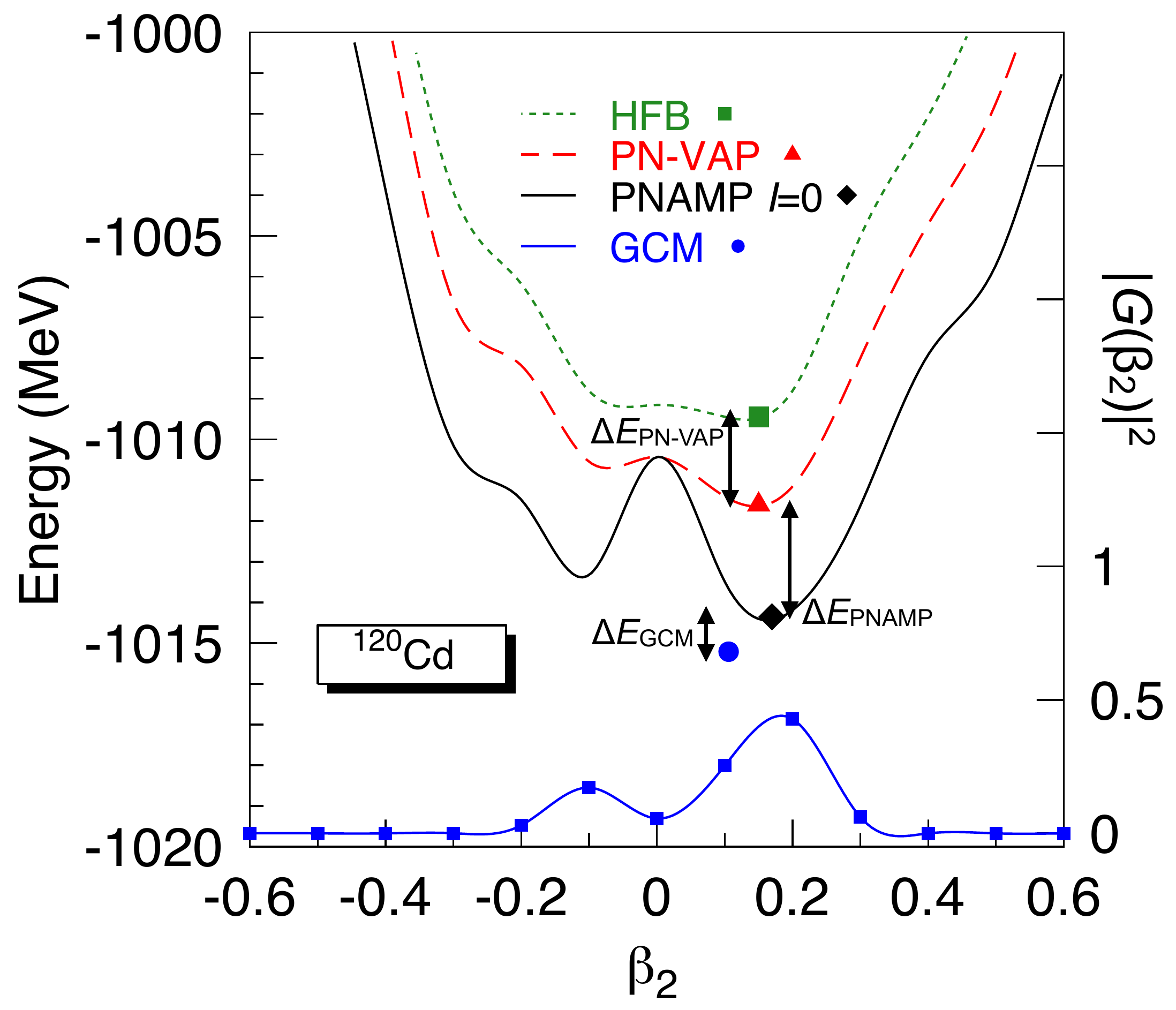}
    \caption{(color online)  Potential energy surfaces as a function of the axial
      quadrupole deformation calculated with HFB (green dotted line),
      PN-VAP (red dashed line) and PNAMP (thin black continuous line)
      approximations for $^{120}$Cd with the Gogny D1S
      parametrization. The square, triangle and diamond represent the
      minima of each surface respectively. The blue dot corresponds to
      the full SCCM energy and the blue boxes (connected by a continuous line to guide the eye) represent the
      ground state collective wave function normalize to 1 ($\sum_{\beta_{2}}|G(\beta_{2})|^{2}=1$). 
      The arrows point out the
      energy gain between the different approaches considered in this
      work. 
     \label{Fig1}}
  \end{center}
\end{figure}

\begin{figure}[tb]
  \begin{center}
    \includegraphics[width=\columnwidth]{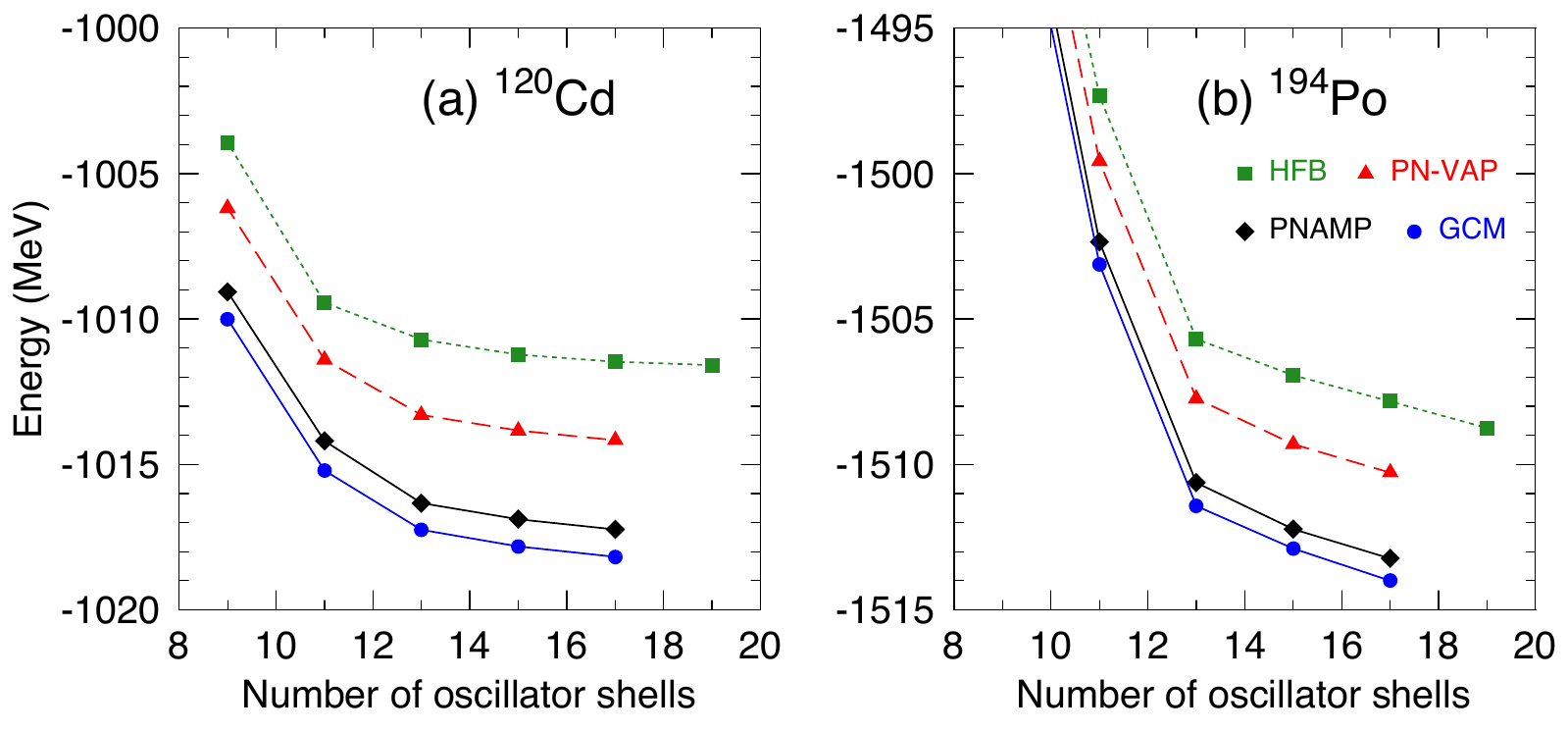}
    \caption{(color online)  Convergence of the energy as a function of the number of major oscillator shells included in the working basis for the same approaches of Fig.~\ref{Fig1}. Left and right panels correspond to $^{120}$Cd and $^{194}$Po respectively, calculated with Gogny D1S.
     \label{Fig_120_194}}
  \end{center}
\end{figure}

\begin{figure}[tb]
  \begin{center}
    \includegraphics[width=\columnwidth]{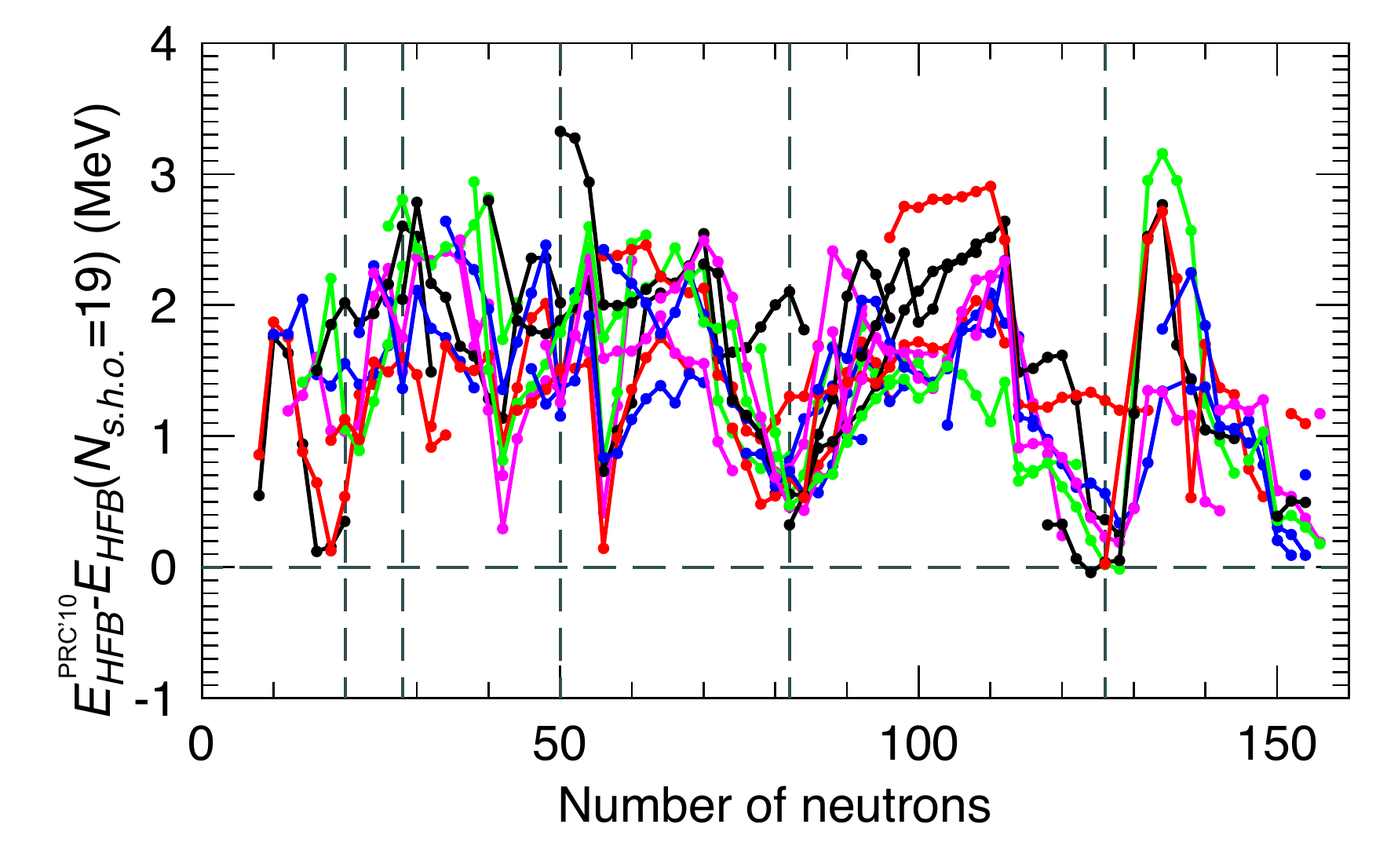}
    \caption{(color online) Differences between the HFB total energies given in Ref.~\cite{PRC_81_014303_2010} and the present HFB calculation performed with $N_{s.h.o.}=19$ and optimized oscillator length. Isotopic chains are connected by lines with the same color and the interaction in all of the cases is Gogny D1S. Positive values mean lower total energies obtained with the present calculation with respect to Ref.~\cite{PRC_81_014303_2010}. \label{D1S_converg}}
  \end{center}
\end{figure}

\begin{figure}[tb]
  \begin{center}
    \includegraphics[width=\columnwidth]{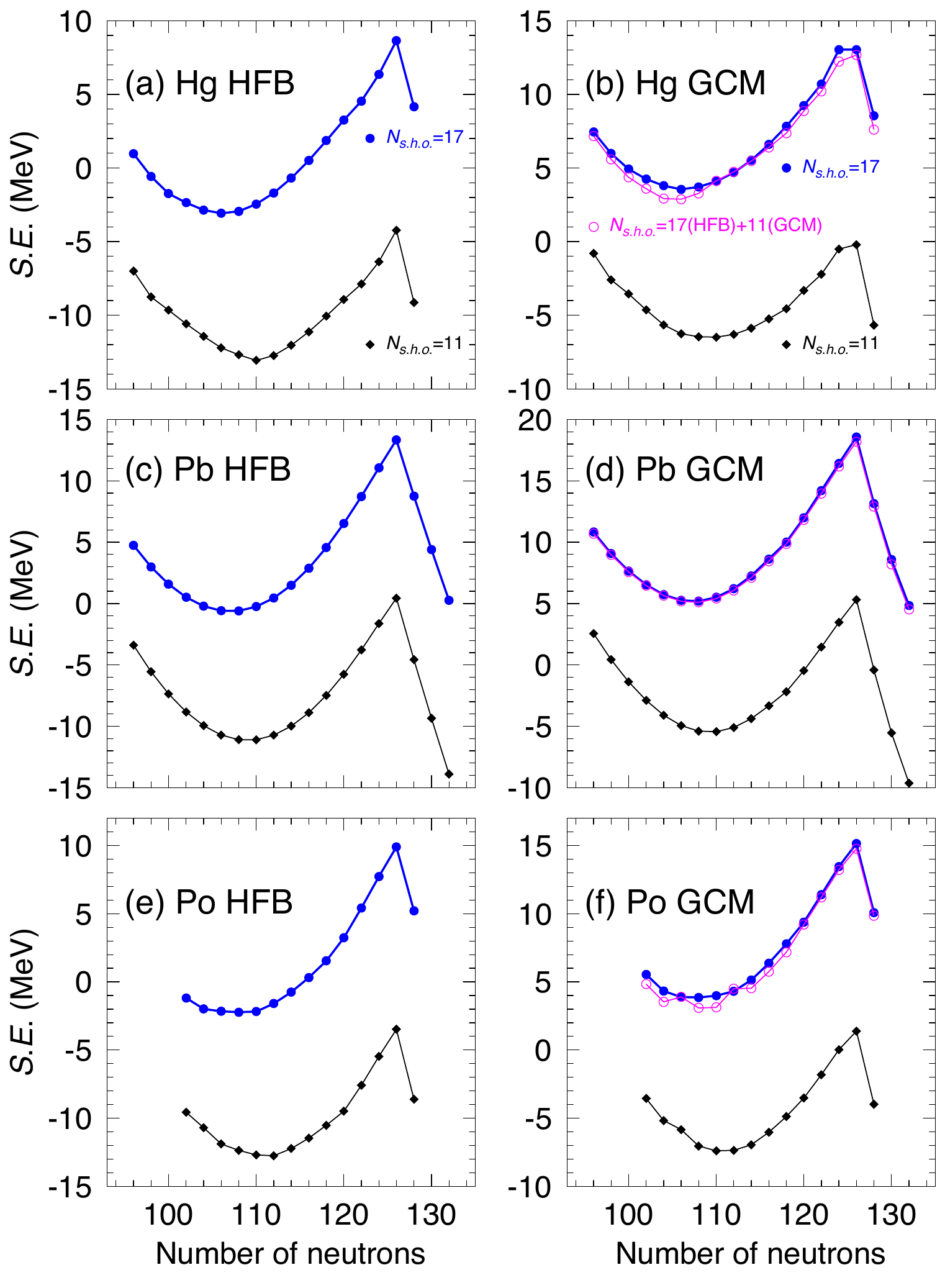}
    \caption{(color online) Shell effects (Eq.~\ref{s.e.}) calculated with the HFB (left panel) and GCM (right panel) methods for Hg (top), Pb (middle) and Po (bottom) isotopic chains. Black diamonds and blue dots are calculated with $N_{s.h.o.}=11$ and 17 respectively. Magenta circles are calculated by adding the BMF corrections computed with $N_{s.h.o.}=11$ to the HFB result with $N_{s.h.o.}=17$. All nuclei are computed with the Gogny D1S parametrization.       
     \label{Fig0}}
  \end{center}
\end{figure}

\begin{figure*}[htb]
  \begin{center}
    \includegraphics*[width=0.9\textwidth]{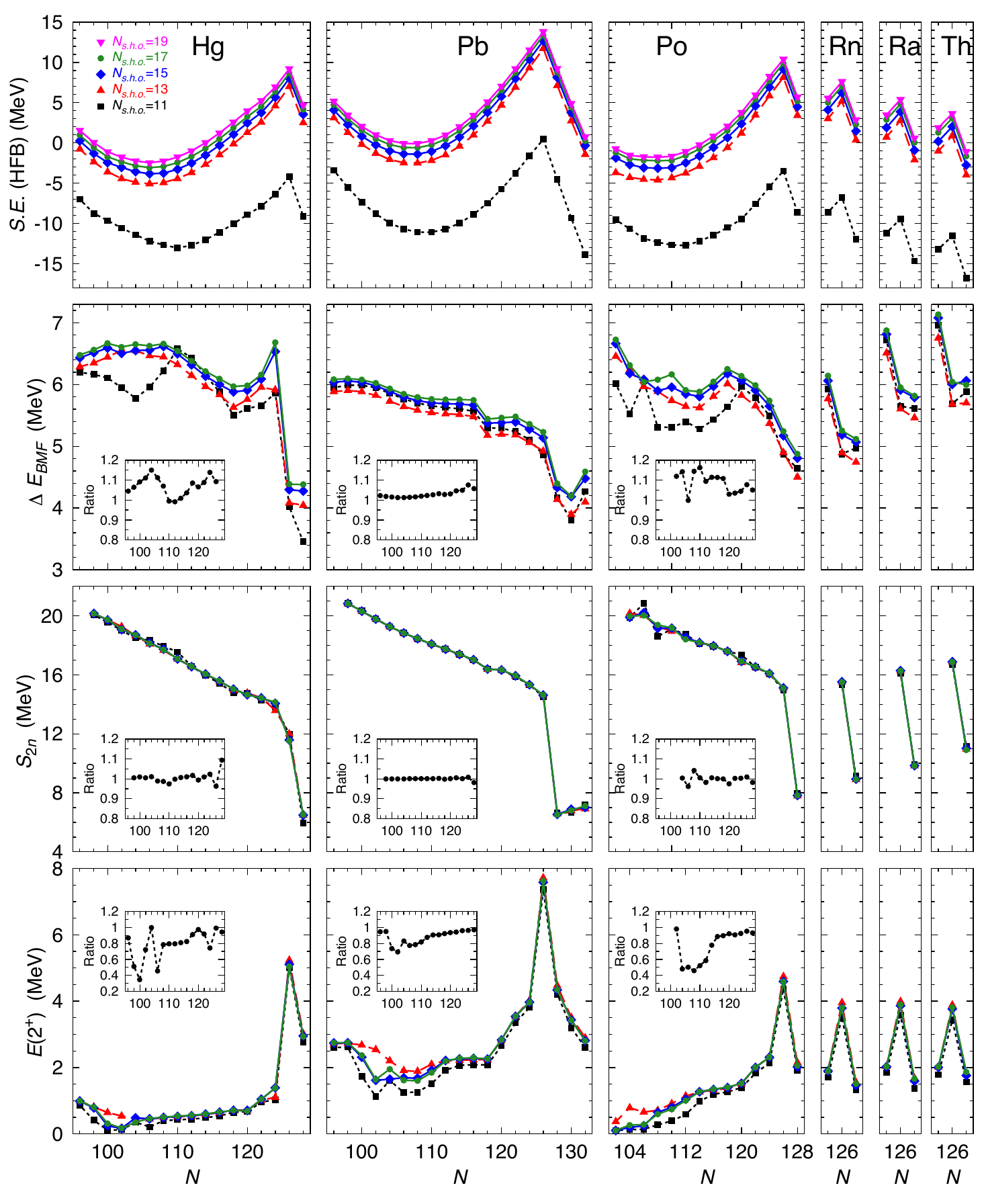}
    \caption{(color online) Results of the calculations performed with different values of $N_{s.h.o.}$ for Hg, Pb, Po, Rn, Ra and Th isotopes (from left to right). Top panel: Shell effects calculated with the HFB method ; top-middle panel: BMF corrections; bottom-middle panel: two-neutron separation energies with BMF corrections; and bottom panel: $2^{+}_{1}$ excitation energies. In the insets, the ratios between the results with $N_{s.h.o.}=17$ and 11 are represented. All nuclei are computed with the Gogny D1S parametrization.\label{Fig_convergence}}
  \end{center}
\end{figure*}

\subsection{Numerical details and convergence of the method}

We summarize some details about the actual global calculations. First of all, 
the HFB~\cite{HFBaxial}, PN-VAP~\cite{NPA_696_467_2001} and SCCM~\cite{PRL_99_062501_2007} codes used throughout this work have been developed in the Nuclear Physics Group at the
Universidad Aut\'onoma de Madrid. The calculations were performed both in GSI-Prometheus (Darmstadt)~\cite{GSI} and CSC-Loewe (Frankfurt)~\cite{CSC} computing facilities, using scripts optimized to perform such a large scale survey. The HFB (mean
field) solutions have been found by using a spherical harmonic
oscillator basis with $N_{s.h.o.}=19$ shells and an optimized
oscillator length for each nucleus~\cite{PRC_21_1568_1980}. Additionally, 
with this large number of s.h.o. shells, potential energy surfaces along the axial quadrupole degree of freedom 
were explored to make sure that the unconstrained HFB-calculations did not converge to local meta-stable 
minima. All terms
(direct, exchange and pairing) in the interaction (including Coulomb)
have been included here and also in the BMF part. For the GCM part, a set
of 15-20 intrinsic many-body wave functions with different axial
quadrupole shapes (oblate and prolate) has been found by using the
PN-VAP method described above. These intrinsic wave functions are
expanded in a basis with $N_{s.h.o.}=11$ shells, again with an
optimized oscillator length for each nucleus. The standard number of
points used in the integrals in the gauge (particle number projection)
and Euler (angular momentum projection) angles were 9 and 16
respectively and the convergence of the quantum number projections
were checked by inspecting the mean values of the operators,
$\hat{N}$, $\hat{Z}$, $\Delta\hat{N}^{2}$, $\Delta\hat{Z}^{2}$ and
$\hat{J}^{2}$.

\begin{table}[b]
\begin{center}
\begin{tabular}{c|c|c|c}\hline \hline
$N_{s.h.o.}$ & HFB & PN-VAP & GCM \\ \hline
11 & $\sim$0.07 h & \bf{$\sim$45 h} & \bf{14.5 h} \\
13 & $\sim$0.20 h & $\sim$120 h & 54.4 h \\
15 & $\sim$0.52 h & $\sim$300 h & 169.1 h \\
17 & $\sim$0.97 h & $\sim$500 h & 460.3 h \\
19 & \bf{$\sim$2.46 h} & -- & -- \\ \hline \hline
\end{tabular}
\end{center}
\caption{Estimation of the computational time used to calculate one nucleus in one single core at the GSI-Prometheus cluster for each level of approximation described in Sec.\ref{Theo}. In boldface, the number of shells chosen in global calculations.}
\label{table_N}
\end{table}%

Finally, the convergence of the solutions of the HWG equations has
been ensured by analyzing the energy plateaus as a function of the
natural states which transform the HWG equations into regular
eigenvalue problems. 
Detailed expressions and performance of this approach can be found in
Refs.~\cite{PRL_99_062501_2007,PRC_81_064323_2010} (and references
therein). 

To shed light on how the BMF method used here actually works, the
nucleus $^{120}$Cd is taken as an example. The HFB --Gogny D1S-- energy
calculated with $N_{s.h.o.}=19$ is $E_{\mathrm{HFB}}=-1011.786$
MeV. On top of this value, BMF corrections are made (see
Eq.~\ref{benergy}). As it is mentioned in the previous section, these
corrections are performed with $N_{s.h.o.}=11$. In Fig.~\ref{Fig1} the
energy as a function of the axial quadrupole deformation $\beta_{2}$
is represented for mean field (HFB, dotted line), variation after
particle number projection (PN-VAP, dashed line) and particle number
and angular momentum $I=0$ projection (PNAMP, $I=0$, thin continuous
line) approximations. The minima of each potential energy surfaces are
the corresponding energies for each level of approximation (square,
HFB; triangle PN-VAP; and diamond PNAMP, $I=0$). In this case, the
value of deformation of all the minima is quite similar
$(\beta_{2}\sim0.17)$, i.e., $^{120}$Cd is prolate deformed in all
of these approaches. A gain in the energy (1.96~MeV) is observed when
correlations associated to the restoration of the particle number are
taken into account. Further correlation energy (2.80~MeV) is obtained
when simultaneous particle number and angular momentum projection is
performed.  In addition, by allowing shape mixing of particle number
and angular momentum restored axial states (GCM),
Eqs.~\ref{GCM_state}-\ref{RitzGCM}, the energy marked by a blue dot in
Fig.~\ref{Fig1} is obtained, i.e., 1.02~MeV extra energy with respect
to the PNAMP ($I=0$) minimum. The square of
so-called ground state collective wave function ($|G(\beta_{2})|^{2}$) is plotted and it
represents the probability of having a given $\beta_{2}$ deformation
in this state (blue boxes in Fig.~\ref{Fig1}). In this case, two maxima are found at $\beta_{2}\sim-0.1$
and +0.2, being the prolate one the absolute maximum. The position of
the blue dot in the abscissa axis corresponds to the mean deformation
calculated with the ground state collective wave function
$\bar{\beta}_{2}=0.10$. In summary, the total correction provided by
the current SCCM method is $\Delta E_{\mathrm{BMF}}=5.77$~MeV and the
total energy (Eq.~\ref{benergy}) is $E(^{120}\mathrm{Cd})=-1017.559$~MeV.

We discuss next the performance and convergence of the results as a function of the number of spherical harmonic oscillator shells, $N_{s.h.o.}$, included in the working basis.
Since the results should not depend on the size of such a basis if a sufficient large number of single particle states are included, in the ideal situation one should take a very large number for $N_{s.h.o.}$. However, this number is limited by the present computational resources.
Thus, the average computing time required at each step of the calculation of one nucleus in a single core,
depending on $N_{s.h.o.}$, is shown in Table~\ref{table_N}. 
Here, we observe the huge differences in the computational burden between the different approaches, in particular when we compare the values needed for HFB and BMF methods. It is important to point out that while the running time for the GCM part can be established beforehand once the number of shells and GCM points are chosen, for the HFB part, and more critically, for the PN-VAP, those numbers can vary from nucleus to nucleus depending on the rate of convergence of the minimization process (Eqs.~\ref{Ritz} and~\ref{RitzPNVAP}). Therefore, the values in Table~\ref{table_N} refer to average numbers in those cases. We can directly check in Table~\ref{table_N} that enlarging the number of shells for BMF calculations would increase prohibitively the computational time. In fact, we have chosen $N_{s.h.o.}=17$ as our current limit for BMF calculations. 

In Fig.~\ref{Fig_120_194} we represent the dependence of the total energy, calculated with the different approaches, on the number of oscillator shells for a medium mass nucleus ($^{120}$Cd mentioned above) and a heavy one ($^{194}$Po). In both cases we observe an energy gain when increasing $N_{s.h.o.}$ but only for $^{120}$Cd a convergence regime is reached for the HFB result (Fig.~\ref{Fig_120_194}(a)). For the heavy nucleus $^{194}$Po (Fig.~\ref{Fig_120_194}(b)), further energy gain is expected if more single particle states are added to the working basis and $N_{s.h.o.}=19$ is not a converged value. Therefore, extrapolation methods to an infinite basis should be applied to further converge the total energy~\cite{NPA_407_1_1983,EPJA_33_237_2007,PRL_102_242501_2009,PRC_86_031301_2012,PRC_87_044326_2013,PRC_89_044301_2014,PRC_90_064007_2014}. The performance and reliability of those extrapolation schemes within the present theoretical framework is a work in progress~\cite{Alex} and we have taken the value with $N_{s.h.o.}=19$ as our best converged one for the HFB result. 

As a matter of fact, in Refs.~\cite{EPJA_33_237_2007,PRC_81_014303_2010} the HFB energy for Gogny D1S has been computed choosing a number of single particle states equal to eight times the larger number among the protons and neutrons in the nucleus. We have checked this prescription by comparing the Gogny-D1S-HFB values given in the supplemental material of Ref.~\cite{PRC_81_014303_2010} with our $N_{s.h.o.}=19$, finding a systematic better convergence in our case of roughly $~1.5$ MeV in the whole nuclear chart (see Fig.~\ref{D1S_converg}). These results point out that convergence of the total energy by using harmonic oscillator bases will be a possible source of problems of the present work which is also shared by previous calculations.

On the other hand, the total energies obtained with PN-VAP, PNAMP and GCM approaches as a function of $N_{s.h.o.}$ are almost parallel to each other, showing that the BMF correlations, $\Delta E_{\mathrm{BMF}}$, are less dependent on the number of single particle states included in the basis (Fig.~\ref{Fig_120_194}). Furthermore, as a consequence of the variational nature of our BMF correlations, larger correlations are obtained with the GCM method than the ones given by PNAMP, being the latter larger than PN-VAP as well.

As it is mentioned above, using Eq.~\ref{benergy} to compute the total energy emerges from the present limitations both in terms of convergence and in computational time. Let us check whether the total energy computed with the SCCM method with a large number of harmonic oscillator shells -$N_{s.h.o.}=17$ is our limit- can be reproduced with the Eq.~\ref{benergy}, computing the HFB part with such a large number of $N_{s.h.o.}$ and $\Delta E_{\mathrm{BMF}}$ with the significantly less time consuming $N_{s.h.o.}=11$ value. 

In Figure~\ref{Fig0} we show the shell effects computed both with HFB and GCM methods for $Z=80$, 82 and 84 isotopic chains. Shell effects are defined as the difference between the total (experimental or theoretical) energy and the energy provided by a liquid drop formula $(E_{LD})$ and it is a convenient way to rescale the total energy:
\begin{equation}
S.E.(Z,N)=E_{LD}(Z,N)-E(Z,N)\label{s.e.}
\end{equation}
We observe first a large difference (up to $\sim10$ MeV) between the shell effects calculated with 11 and 17 oscillator shells within the same many-body approach, HFB or GCM, for all isotopic chains. This shows again explicitly that the total energy calculated with $N_{s.h.o.}=11$ is not well converged, now in three different isotopic chains. However, we can approach the BMF results with $N_{s.h.o.}=17$ by adding to the HFB values computed with the same number of oscillator shells, the BMF corrections obtained with $N_{s.h.o.}=11$. Such a result is represented with magenta circles in Figs.~\ref{Fig0}(b)-(d)-(f), showing a nice agreement with the GCM values calculated the largest value of $N_{s.h.o.}$ that we can reach for this approach.
\begin{figure}[b]
  \begin{center}
    \includegraphics[width=0.9\columnwidth]{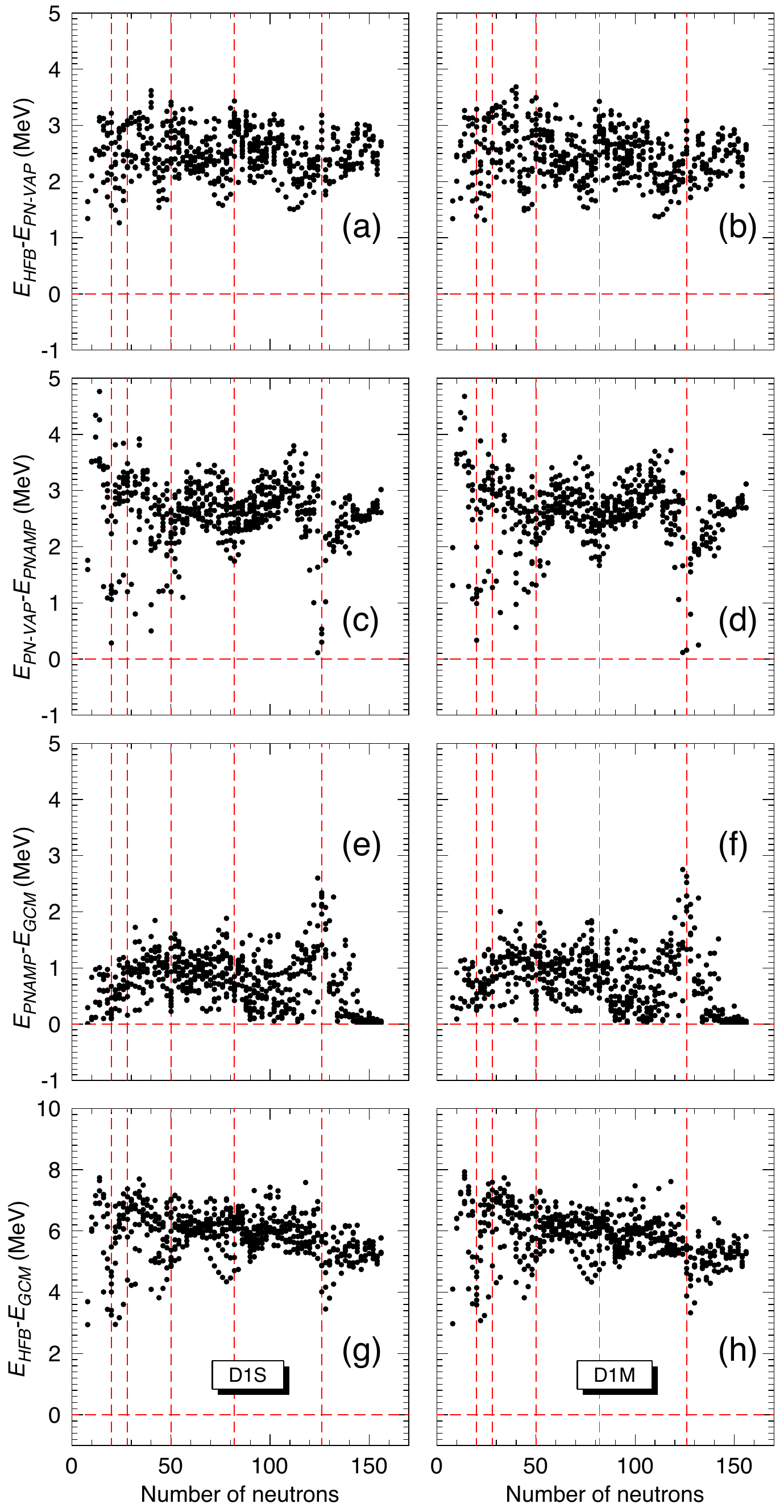}
    \caption{(color online) Gain in total energy as a function of the
      number of neutrons obtained by including (a)-(b) variation after
      particle number projection (PN-VAP) method; (c)-(d) Simultaneous
      particle number and angular momentum $I=0$ projection (PNAMP);
      (e)-(f) quadrupole shape mixing within the generator coordinate
      method (GCM) with symmetry restored states; and (g)-(h) total
      beyond-mean-field gain. Dashed vertical lines represent the
      neutron magic numbers 20, 28, 50, 82 and 126. Left and right
      panels correspond to Gogny D1S and D1M parametrizations,
      respectively. \label{corr_N}}
  \end{center}
\end{figure}

Once we have checked the suitability of splitting the total energy in a MF part plus BMF corrections, we have performed additional tests of the convergence of the results in heavy nuclei. Hence, $Z=80-84$ isotopic chains and $N=124-128$ isotonic chains have been calculated in all of the approaches with increasing values of the oscillator shells (with the Gogny D1S parametrization). The results are summarized in Fig.~\ref{Fig_convergence} where shell effects (top panel), beyond mean field corrections (top-middle panel), two-neutron separation energies (bottom-middle panel) and $2^+$ excitation energies (bottom panel) are shown.
In the top panel of Fig.~\ref{Fig_convergence} the shell effects corresponding to HFB calculations are represented, although similar patterns but shifted to larger values are found in PN-VAP and GCM results (see right panel of Fig.~\ref{Fig0}). We observe that increasing $N_{s.h.o.}$ the results tend to collapse to a final curve. However, the differences between the HFB results for $N_{s.h.o.}=19$ and 17 are around $\sim0.8$ MeV in average and one should go to larger number of oscillator shells to get full convergence. 

The large differences found in the shell effects for the $N_{s.h.o.}=11$ and $N_{s.h.o.}=19$ results (up to 14 MeV) are much smaller in the calculated BMF corrections. Furthermore, they are almost negligible in the  two-neutron(proton) separation energies -defined in Eq.~\ref{s2n}. However, we also observe some local deviations from the $N_{s.h.o.}=17$ - chosen to be the best values for BMF approaches- in the neutron deficient Hg, Pb and Po isotopes. In this region, several deformed configurations are almost degenerated (see Ref. ~\cite{PRC_89_014306_2014} and references therein) that can be favored differently depending on the number of oscillator shells. Hence, those small jumps are produced by a change in deformation. In any case, the largest difference are around 0.9 MeV but still could lead to small artificial jumps in $S_{2n(2p)}$ and/or $E(2^{+}_{1})$. The above aspects are also visualized in the corresponding ratios between the results with $N_{s.h.o.}=17$ and $N_{s.h.o.}=11$ represented in the insets of Fig.~\ref{Fig_convergence}.  Therefore, these convergence effects should be taken into account for improving the precision of the mass models.  

Finally, the comparison with the available experimental data for even-even nuclei requires the calculation of 598 nuclei for each parametrization of the Gogny functional, i.e., 1196 nuclei. Therefore, as it was mentioned in previous sections, the present global survey has been restricted to compute the HFB energy and the beyond-mean-field corrections using $N_{s.h.o.}=19$ and $11$ respectively. According to the previous analysis, this choice seems to be a reasonable compromise between convergence and computational time. This means that the calculation of one nucleus within the prescription followed in this work takes $\sim62$ hours/core at the GSI-Prometheus cluster, which is still feasible with our current facilities.

\section{Results}
\label{Res}

\subsection{BMF correlation energies}\label{corr_ener}

We now generalize the results obtained in Fig.~\ref{Fig1} to the even-even nuclei with $Z,N\geq
10$ contained in the most recent Atomic Mass Evaluation
(AME)~\cite{AW} both for Gogny D1S and D1M parametrizations. In
Fig.~\ref{corr_N} the successive gains in total energy reached by
restoring the symmetries and allowing the axial quadrupole shape
mixing are represented as a function of the number of neutrons. 
The first noticeable aspect is the striking similarity between those correlation energies
for both D1S and D1M parametrizations. 
We observe a band of values ranged in the
interval 1.5--3.5~MeV with a mean gain $\sim2.3$~MeV of correlation
energy with respect to the mean field (HFB) solutions when variation
after particle number projection (PN-VAP) method is applied
(Figs.~\ref{corr_N}(a)-(b)). In addition, some
local minima are obtained around $\sim24, \sim44, \sim78$ and
$\sim110$, right before the neutron magic numbers $28, 50, 82$ and
$126$ for both D1S and D1M parametrizations.

Minima are also found both in the energy gained by particle number and
angular momentum restoration (PNAMP) -on top of PN-VAP- and by the
generator coordinate method (GCM) -on top of PNAMP- but now located at the shell
closures. In the former (Figs.~\ref{corr_N}(c)-(d)), larger
correlation energies are obtained in the mid-shell and minima are
found at the neutron magic numbers. Excluding the lighter and the
semi-magic nuclei, an average gain of $\sim 2.7$ MeV is attained and a
slightly decreasing slope is also observed when increasing the number
of neutrons. Concerning the GCM correlation energies
(Figs.~\ref{corr_N}(e)-(f)), contrary to PNAMP, the larger gains are
almost at the shell closures, obtaining a clear maximum at
$N=126$. The average gain in this case is $\sim0.8$ MeV. This
behavior of the BMF energies is important to correct the parabolic
shape observed in experimental theoretical energy differences at the
HFB level (see fig.~\ref{Masses_AW}). 

Finally, the total beyond-mean-field energy gain (summing up all of
the previous contributions) is represented in
Figs.~\ref{corr_N}(g)-(h). We observe first that smaller correlation
energies are obtained around the magic numbers, producing
qualitatively an inverted parabolic behavior between two consecutive
shell closures. Furthermore, we see a larger
spread in the BMF energy gain in the lighter nuclei. Finally, the
overall gain is slightly smaller when the
number of neutrons is increased for both parametrizations. 

\begin{figure}[htb]
  \begin{center}
    \includegraphics[width=0.9\columnwidth]{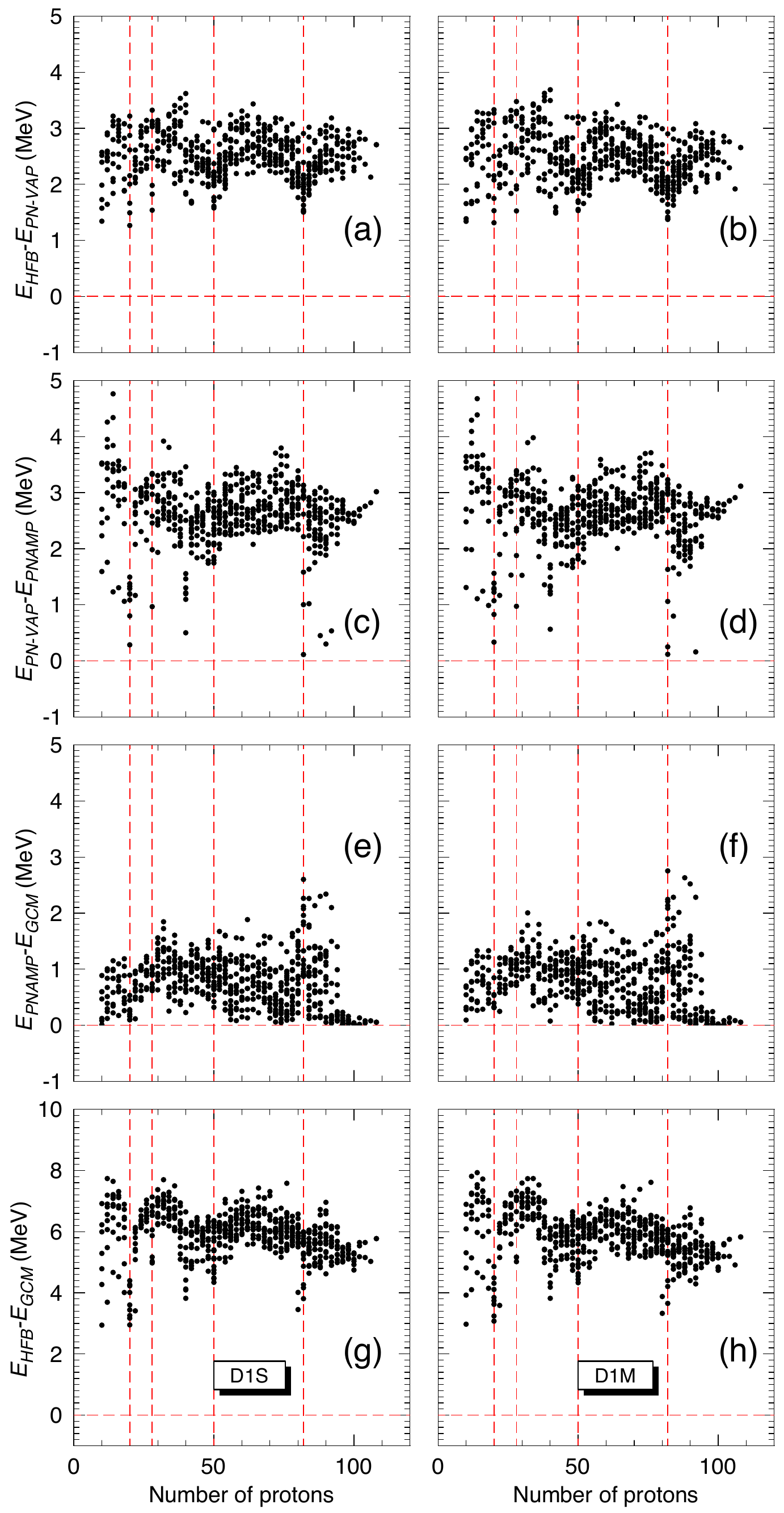} 
    \caption{Same as Fig.~\ref{corr_N} but as a function of the number of
      protons. Dashed vertical lines represent the proton magic numbers
      20, 28, 50 and 82.\label{corr_Z}}
  \end{center}
\end{figure}

Analogous patterns as the ones just described above are found for the
different levels of approximation when they are represented as a
function of the number of protons (see Fig.~\ref{corr_Z}). Hence, a
rather flat energy gain for the PN-VAP approach is obtained both for D1S and 
D1M parametrizations (Figs.~\ref{corr_Z}(a)-(b)). Furthermore, the local minima are found now at the proton
magic numbers 20, 28, 50 and 82 in this approximation. For PNAMP (Figs.~\ref{corr_Z}(c)-(d)) and
GCM (Figs.~\ref{corr_Z}(e)-(f)) approaches we do not observe differences either between the
parametrizations. In the former, shell effects are still present at
$Z=$ 20, 28, 40 and 82; in the latter, the lead isotopes have the most
prominent ones.  

\begin{figure}[t]
\begin{center}
\includegraphics[width=0.9\columnwidth]{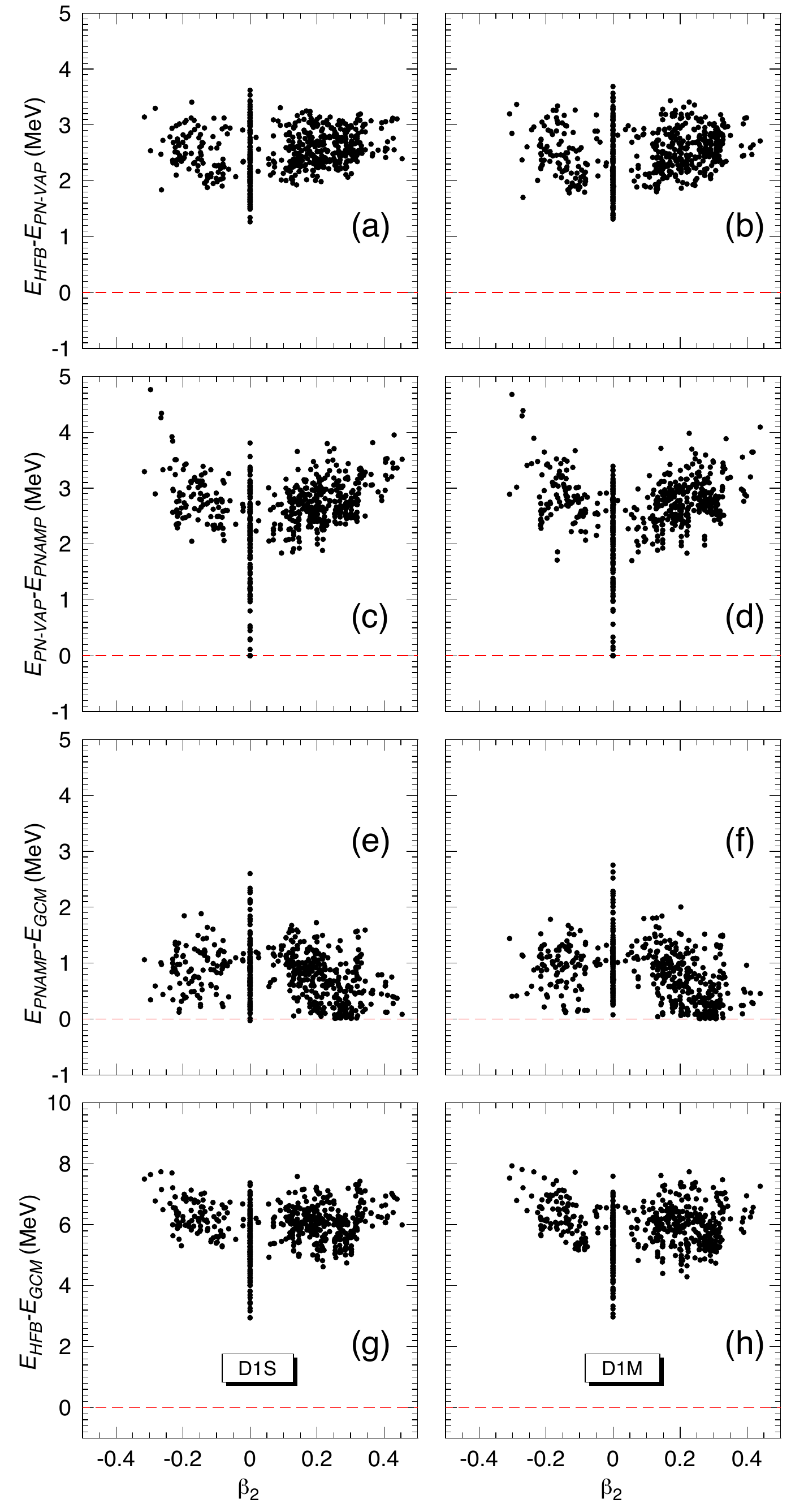}
\caption{Same as Fig.~\ref{corr_N} but as a function of the quadrupole deformation found at the mean field -HFB- level.}
\label{corr_def}
\end{center}
\end{figure}

\begin{figure}[tb]
  \begin{center}
    \includegraphics[width=\columnwidth]{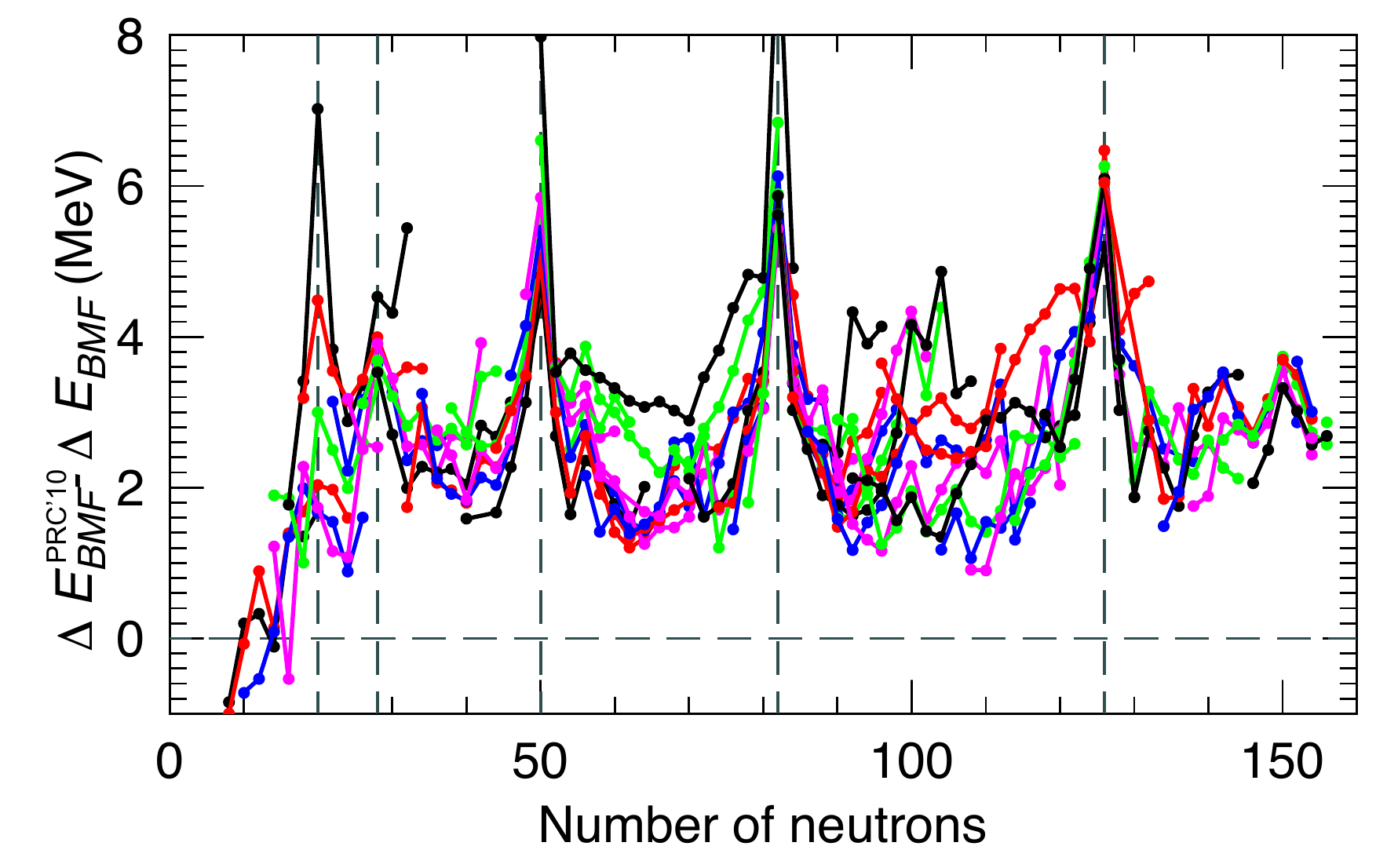}
    \caption{(color online) Difference between the total correlation energies given by the 5DCH model in Ref.~\cite{PRC_81_014303_2010} and the ones calculated with the present SCCM method. Isotopic chains are connected by lines with the same color and the interaction in all of the cases is Gogny D1S.     
     \label{D1S_5DCH}}
  \end{center}
\end{figure}

Since both PNAMP and GCM methods depend crucially on the intrinsic
deformation, we also represent the correlation energy gains obtained
for the different approaches as a function of the quadrupole
deformation $\beta_{2}$ obtained at the HFB level for each nucleus
(Fig.~\ref{corr_def}). In all cases, the largest spread in correlation
energy gain is obtained in the spherical point $\beta_{2}=0$. Hence,
for this intrinsic deformation we observe energy gains ranging from 1.2-3.5~MeV,
0.0-3.7~MeV and 0.0-2.8~MeV for PN-VAP (Fig.~\ref{corr_def}(a)-(b)), PNAMP (Fig.~\ref{corr_def}(c)-(d)), and
GCM (Fig.~\ref{corr_def}(e)-(f)), respectively. These spreads are smaller for the rest of
deformations. Additionally, the energy gain does not depend very much
on the size and sign of the quadrupole deformation both in the PN-VAP (Fig.~\ref{corr_def}(a)-(b))
and GCM (Fig.~\ref{corr_def}(e)-(f)) approaches. For PNAMP case (Fig.~\ref{corr_def}(c)-(d)), a mild trend of having
larger energy gains with increasing $|\beta_{2}|$-values is
obtained. Similar results are obtained with Skyrme functionals (see
Fig. 7 of Ref~\cite{PRC_73_034322_2006}) showing that these patterns
depend on the method rather than on the choice of the functional. 

 To end this section, we also compare the total correlation
  energies obtained with the present axial SCCM calculations with
  those provided by the 5DCH for Gogny D1S given in the supplemental
  material of Ref.~\cite{PRC_81_014303_2010}. We observe in
  Fig.~\ref{D1S_5DCH} that, except for a few light nuclei, the
  correlations given by the present method are about $2-3$ MeV larger
  than the 5DCH ones. Even larger differences are found around the
  shell closures since the 5DCH calculation gives positive correlation
  energies in those nuclei. As a matter of fact, such a
  anti-correlation energy are set to zero in
  Ref.~\cite{PRC_81_014303_2010}. Since the 5DCH method is based on a
  gaussian overlap approximation, then it is not variationally
  consistent with the underlying HFB functional. Furthermore, although the triaxial degree of freedom
  is included in the collective hamiltonian, the present axial SCCM
  calculations include both symmetry restorations and GCM without GOA
  that produce larger correlation energies. These SCCM correlation
  energies are negative everywhere (see Fig.~\ref{corr_N}(g) for
  instance) and including other degrees of freedom such as the
  triaxiality will produce even more negative values. 
    In addition, although convergence of both HFB and SCCM is not
    globally reached we can at least determine where it has been
    achieved. Fig.~\ref{D1S_5DCH} shows that the difference in
    correlation energies between the present calculations and the 5DCH
    approach is rather constant along the whole nuclear chart, apart
    from the spikes around shell closures. This indicates that the 5DCH
    approach may suffer from similar convergence issues for heavy
    nuclei. However, one should keep in mind that due to its
    non-variational nature one cannot strictly speak of converged 5DCH
    calculations.

\subsection{Comparison with experimental data}

\begin{figure*}[htb]
  \begin{center}
    \includegraphics[width=0.8\textwidth]{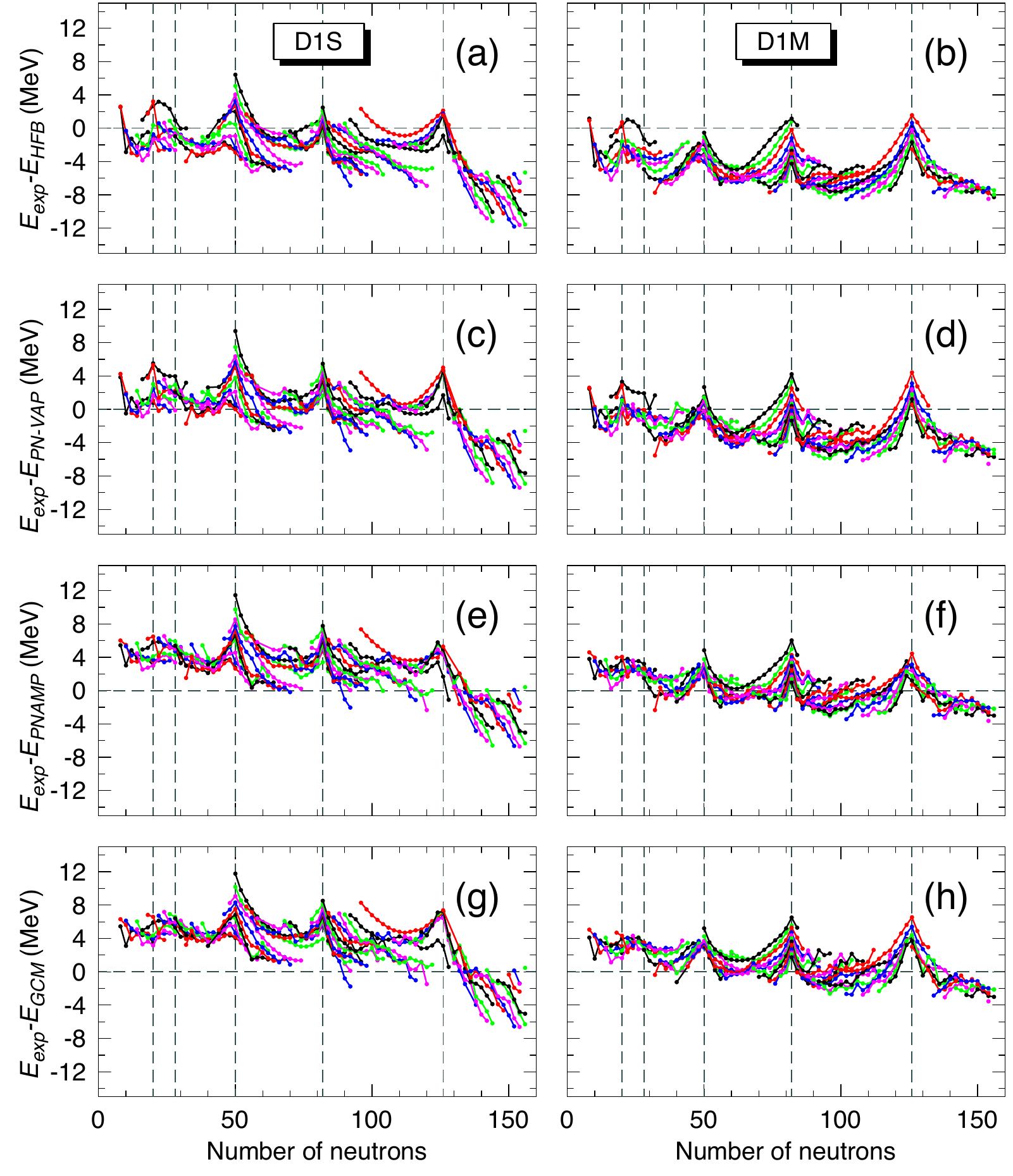} 
    \caption{(color online) Difference between the experimental
      total energies (taken from Ref.~\cite{AW}) and: (a)-(b) HFB;
      (c)-(d) PN-VAP; (e)-(f) PNAMP; (g)-(h) GCM total energies
      calculated with the Gogny D1S (left panel) and D1M (right panel)
      parametrizations. Lines connect isotopic chains starting from
      $Z=10$. Black, red, blue, magenta and green lines represent
      isotopic chains with $Z= x0, x2, x4, x6$ and $x8$, being
      $x=1,2,...$, etc. Dashed vertical lines mark the neutron magic
      numbers 20, 28, 50, 82 and 126.\label{Masses_AW}}
  \end{center}
\end{figure*}

\begin{figure*}[htb]
  \begin{center}
    \includegraphics[width=0.8\textwidth]{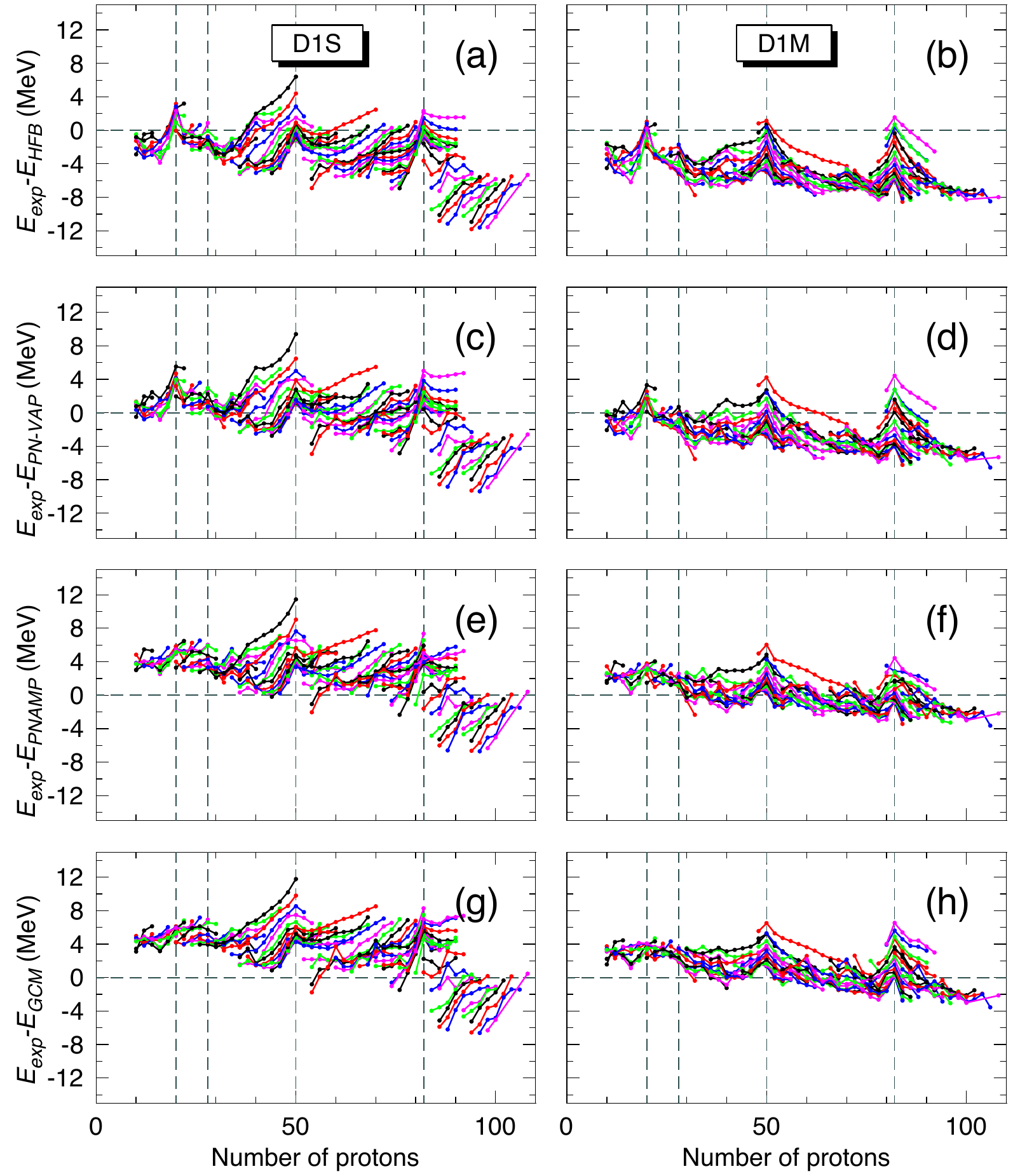}
    \caption{(color online) Same as Fig.~\ref{Masses_AW} but as a
      function of the number of protons. Lines connect isotonic chains
      starting from $N=10$. Black, red, blue, magenta and green lines
      represent isotopic chains with $N= x0, x2, x4, x6$ and $x8$,
      being $x=1,2,...$, etc. Dashed vertical lines mark the neutron
      magic numbers 20, 28, 50 and 82.\label{Masses_AW_ZZ}}
  \end{center}
\end{figure*}

\begin{figure}[htb]
  \begin{center}
    \includegraphics[width=0.7\columnwidth]{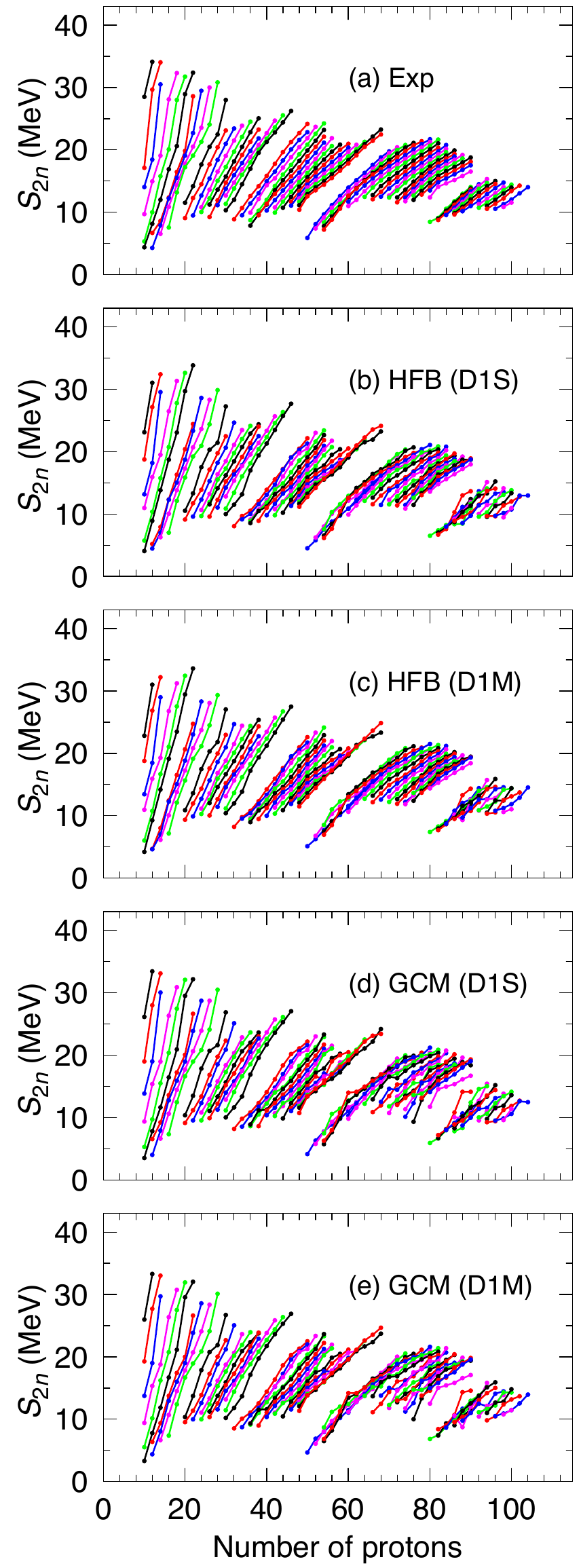}
    \caption{(color online) Two-neutron separation energies as a
      function of the number of protons: (a) Experimental data (taken
      from Ref.~\cite{AW}); (b) HFB; (d) GCM calculated with the Gogny
      D1S parametrization and (c) HFB ; (e) GCM calculated with D1M
      parametrization. Lines connect isotonic chains starting from
      $N=10$. Black, red, blue, magenta and green lines represent
      isotopic chains with $N= x0, x2, x4, x6$ and $x8$, being
      $x=1,2,...$, etc.\label{S2n_AUDI}}
  \end{center}
\end{figure}

\begin{figure}[htb]
  \begin{center}
    \includegraphics[width=0.7\columnwidth]{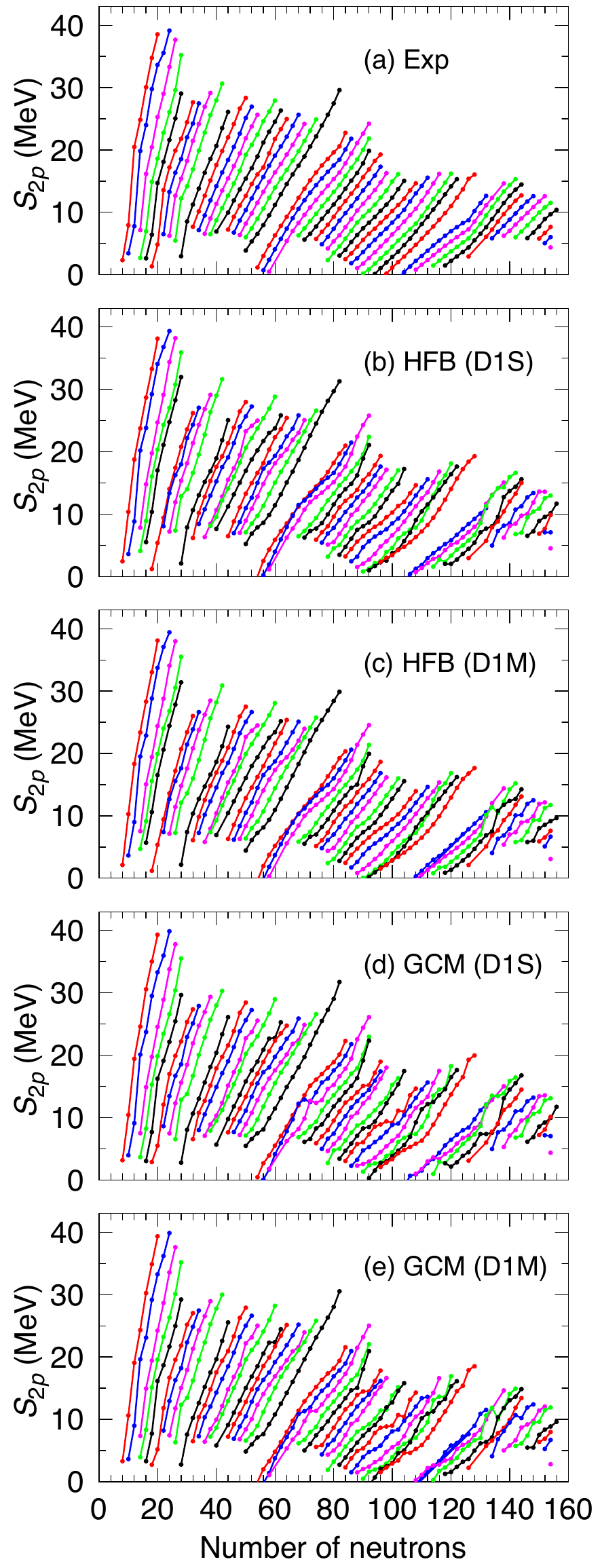}
    \caption{(color online) Two-proton separation energies as a
      function of the number of neutrons: (a) Experimental data (taken
      from Ref.~\cite{AW}); (b) HFB; (c) GCM calculated with the Gogny
      D1S parametrization. Lines connect isotopic chains starting from
      $Z=12$. Black, red, blue, magenta and green lines represent
      isotopic chains with $Z= x0, x2, x4, x6$ and $x8$, being
      $x=1,2,...$, etc...\label{S2p_AUDI}} 
  \end{center}
\end{figure}

\subsubsection{Masses}

We now compare the results obtained with the different approximations
with the experimental data extracted from the most recent
AME~\cite{AW}. 
In Fig.~\ref{Masses_AW} the differences between the experimental and
theoretical masses are plotted for the different mean field and beyond
mean field approaches. More quantitative results are written in
Table~\ref{Tab1} for the D1S and D1M parametrizations.  

Starting from the oldest parametrization, i.e., D1S, we notice first
its poor performance in describing experimental
masses. In none of the many-body approaches studied here, the root
mean square (RMS) deviation is less than 2.6 MeV. This is explained by
three major drawbacks of this parametrization (see left panel of
Fig.~\ref{Masses_AW}). The first one is the presence of residual shell
effects. In all of the approaches, peaks at the neutron magic numbers
$N=50$, 82 and 126 are observed. As it was discussed above, BMF energy
gains are smaller in the shell closure nuclei (see
Fig.~\ref{corr_N}(g)). Therefore, these peaks are reduced when BMF
effects are taken into account but the reduction is clearly
insufficient to bring the theory closer to the experiment. The second
drawback is the systematic drift towards less bound systems in nuclei
with increasing neutron excess. The origin of the problem is in the
symmetry energy provided by Gogny D1S. This parametrization does not
reproduce the correct curvature in the neutron matter equation of
state given by \textit{ab-initio} approaches~\cite{PLB_668_420_2008}
producing a lack of binding energy in neutron rich nuclei. Again, BMF
effects do not change this trend.  Nevertheless, the spread in light
nuclei (from $N=10-40$) found at the mean field and PN-VAP
approximations (Fig.~\ref{Masses_AW}(a)) is significantly reduced when
PNAMP and GCM are taken into account (Figs.~\ref{Masses_AW}(e)-(g)).
The third drawback is the way in which the parameters of the
interaction were obtained. Hence, the parameters of the oldest
realizations of the Gogny interaction were fitted to reproduce
experimental data with the HFB method but leaving some room for
eventual beyond-mean-field effects~\cite{PRC_21_1568_1980}. However,
some overbinding is still obtained with respect to the experimental
values. The evolution of the RMS values given in the
second column of Tab.~\ref{Tab1} reflects also this effect, obtaining
for the most sophisticated many-body method used in this work a RMS
deviation of 4.45 MeV (for 598 masses).

The D1M parametrization~\cite{PRL_102_242501_2009} was built to
correct these shortcomings of the D1S by performing a fit to a large
set of experimental masses using the 5DCH
method~\cite{PRC_81_014303_2010} to include beyond-mean-field effects. 
That led to a RMS deviation from data of $\sim0.798$~MeV
(for 2149 masses).  Except for the inclusion of triaxiality and the
lack of quantum number projections, the 5DCH method can be considered
as a gaussian overlap approximation (GOA) of the method used in this
work~\cite{RING_SCHUCK}. Let us analyze now the performance of D1M in
combination with the present axial SCCM method which does not assume such a GOA approximation. In the
right panel of Fig.~\ref{Masses_AW} the difference between experimental and theoretical masses
obtained with the D1M parametrization are shown. Here we observe that
the drift and, partially, the overbinding found with the D1S
parametrization are corrected. However, as it was already stated in
Ref.~\cite{PRL_102_242501_2009}, strong shell effects are still
present and the theoretical results that overestimate the binding
energies around the magic neutron numbers, particularly at $N=50$, 82
and 126. This behavior is not corrected by including BMF correlations
of the kind studied in this work. Nevertheless, the addition of
correlations improves the agreement with data with respect to the mean
field results. Since the D1M parameters were fitted taking already into
account BMF effects, the results at the HFB level are
underbound with respect to the experimental values
(Fig.~\ref{Masses_AW}(b)). A very large RMS deviation is obtained for
this approach and a much smaller for the rest (see
Table~\ref{Tab1}). However, the correlation energies attained by the
GCM are larger than the ones provided by the 5DCH, as discussed above. This produces an
excess of total energy also with this parametrization when the axial
shape mixing with quantum number projection is taken
into account (Fig.~\ref{Masses_AW}(h)).  The RMS value for the
GCM approach with the D1M parametrization is 2.17 (for 598
masses).

 It is worth mentioning that the global RMS values themselves
  do not provide the complete picture of the comparison with the
  experimental data since some compensation effects can occur in
  different regions of the nuclear chart.  A RMS-value independent on the number of particles 
   -including BMF effects- must
  be pursued even though an excess of total energy would be expected
  when BMF corrections are incorporated on top of a functional fitted
  to reproduce data with less correlated ground states. In this sense,
  BMF corrections work better in reducing the spread and shell gaps in
  $N\leq28$ nuclei than in heavier systems (see Fig.~\ref{Masses_AW})
  although, from the point of view of the RMS, light nuclei are
  farther away from the experiment than the heavy ones.

\begin{table}[htb]
  \caption{\label{Tab1} RMS comparison between theoretical calculations
    and experimental data for total energies, two neutron and two
    proton separation energies. All energies are in MeV.}
  \begin{ruledtabular}
    \begin{tabular}{cccccc}
D1S & $E$  & $S_{2n}$ & $S_{2p}$  \\
\hline
HFB 	& 3.53 & 0.98 & 1.15  \\
PN-VAP & 2.62 & 1.10 & 1.11  \\
PNAMP & 3.75 & 0.98 & 1.00  \\
GCM	 & 4.45 & 0.95 & 1.00  \\
\hline
D1M & $E$  & $S_{2n}$ & $S_{2p}$  \\
\hline
HFB 	& 5.29 & 0.89 & 0.99	 \\
PN-VAP & 3.14	& 1.03 & 0.96  \\
PNAMP & 1.79	& 0.89 & 0.86  \\
GCM	 & 2.17 & 0.85 & 0.87  \\
\end{tabular}
\end{ruledtabular}
\end{table}%

\begin{table}[htb]
\caption{\label{Tab2} Logarithmic errors and deviations for the $2^{+}_{1}$ excitation energies computed for GCM-D1S, GCM-D1M, GCM-SLy4~\cite{PRC_75_044305_2007} and 5DCH-D1S~\cite{PRL_99_032502_2007,PRC_81_014303_2010}. Experimental data are taken from Ref.~\cite{ADNDT_78_1_2001}.}
\begin{ruledtabular}
\begin{tabular}{ccccccc}
 & GCM-D1S  & GCM-D1M & GCM-SLy4 & 5DCH-D1S  &  5DCH-D1S \\
 & (534) & (534) & (359) & (513) & (519) \\
\hline
$\bar{R}_{E}$ & 0.32 & 0.35 & 0.51 & 0.11 & 0.12 \\
$\sigma_{E}$ & 0.42 & 0.43 & 0.38 &  0.35 & 0.33 \\
\end{tabular}
\end{ruledtabular}
\end{table}%

\begin{figure*}[htb]
  \begin{center}
    \includegraphics[width=1\textwidth]{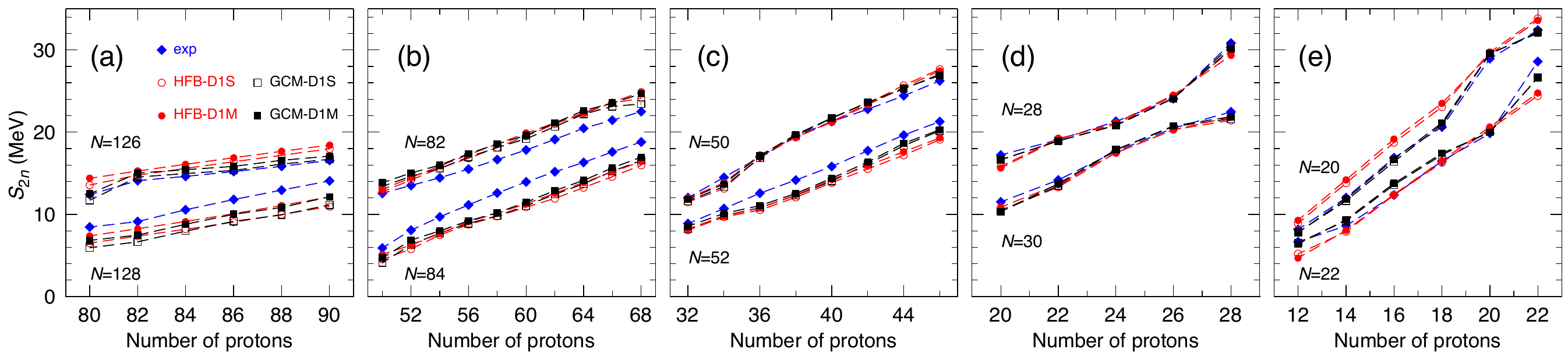}
    \caption{(color online) Two-neutron separation energies for (a) $N=126,128$; (b) $N=82,84$; (c) $N=50,52$; (d) $N=28,30$; and (e) $N=20,22$ isotonic chains. Blue diamonds, red circles and black boxes represent the experimental, mean field and GCM results respectively. Empty (full) symbols are calculated with the Gogny D1S (D1M) parametrization.  \label{shell_gaps_neutrons}}
  \end{center}
\end{figure*}
\begin{figure*}[htb]
  \begin{center}
    \includegraphics[width=0.8\textwidth]{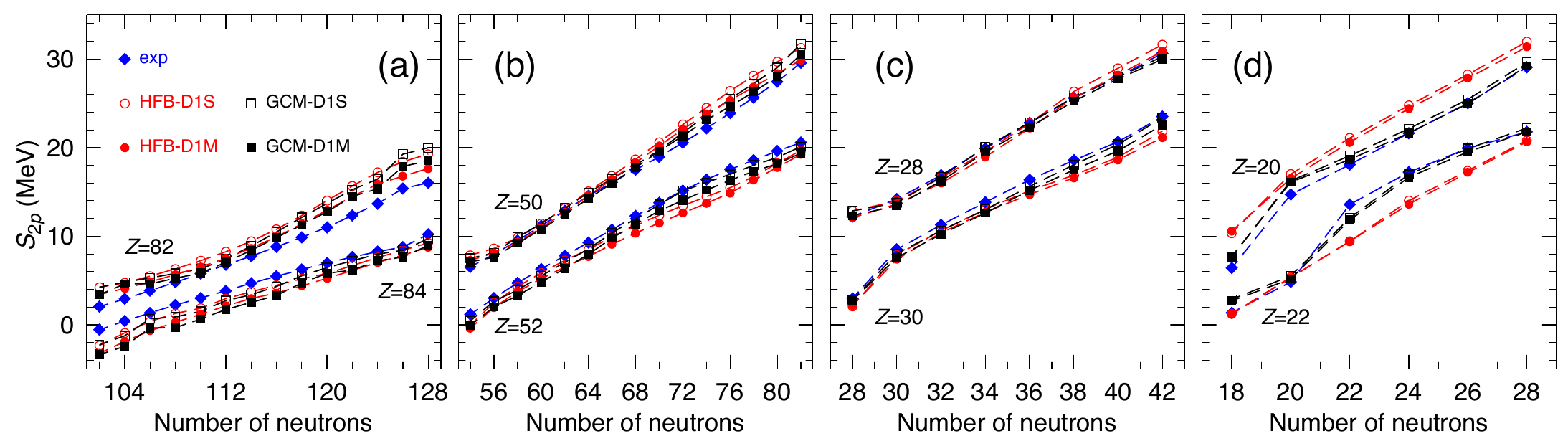}
    \caption{(color online) Two-proton separation energies for (a) $Z=82,84$; (b) $Z=50,52$; (c) $Z=28,30$; and (d) $N=20,22$ isotopic chains. Blue diamonds, red circles and black boxes represent the experimental, mean field and GCM results respectively. Empty (full) symbols are calculated with the Gogny D1S (D1M) parametrization.  \label{shell_gaps_protons}}
  \end{center}
\end{figure*}

\begin{figure*}[htb]
  \begin{center}
    \includegraphics[width=0.8\textwidth]{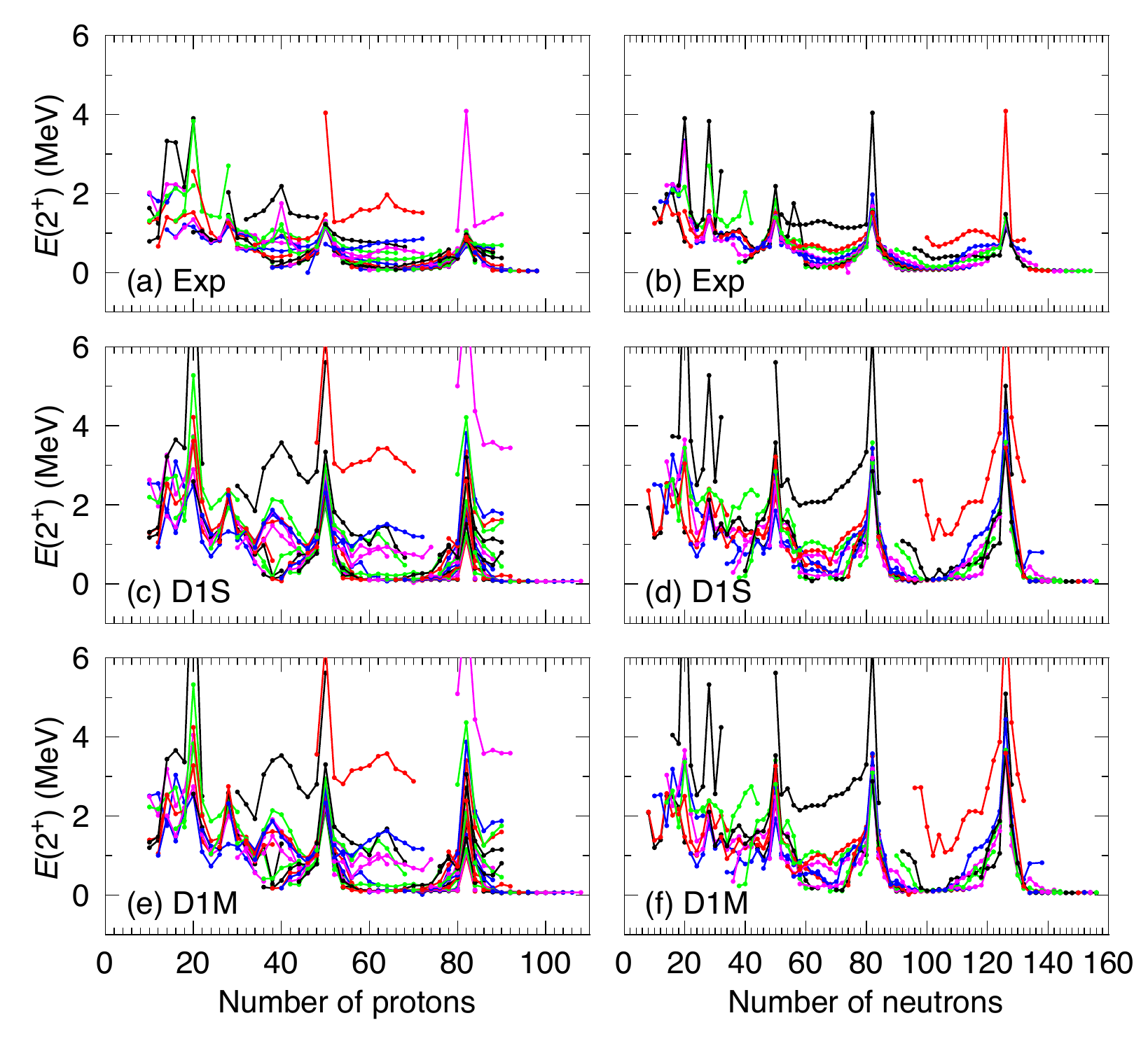}
    \caption{(color online) $2^{+}_{1}$ excitation energies for:
      (a)-(b) Experimental data (taken from
      Ref.~\cite{ADNDT_78_1_2001}); (c)-(d) GCM-Gogny D1S; and (e)-(f)
      GCM-Gogny D1M. Lines connect isotonic chains starting from
      $N=10$ (left panel) and isotopic chains starting from $Z=10$
      (right panels). The color code is the same as in previous
      figures.\label{e2_d1s_d1m}}
  \end{center}
\end{figure*}

In order to check the possible shells effects appearing in isotonic
chains, we represent in Fig.~\ref{Masses_AW_ZZ} the energy differences with 
the experimental masses obtained
for the different many-body approaches and parametrizations as a
function of the number of protons.  These shell effects, though still
present, are slightly less pronounced than in the isotopic
chains. This is in agreement with the results found with Skyrme
interactions~\cite{PRC_73_034322_2006}. However, it is interesting to
note that the relative overbinding found at $Z=20$ at the HFB level is
smoothened out with BMF approaches in both parametrizations while
$Z=50$ and 82 persist. Finally, we observe a clear difference between
the results provided by the two parametrizations. For D1M rather
symmetric energy differences around the shell closures are found while for D1S
larger energies are obtained for larger values of $Z$ within a given
isotonic chain. Since the isotonic chains start normally with $N>Z$,
this behavior reflects again a lack of ground state energy provided by
the D1S parametrization in nuclei with neutron excess, i.e., the
symmetry energy problem already mentioned.

To conclude this section, we can state that both the symmetry energy
problem and the overbinding produced by the inclusion of BMF effects
can be solved by modifying the parametrization, as it is almost done
with the introduction of the Gogny D1M interaction. However, the
energy excess obtained in the magic nuclei (relative to the energy
predicted in the mid-shell nuclei), although reduced, is not washed
out by taking into account the present BMF effects. It is still an
open question whether the current BMF functionals, with parameters
self-consistently fitted and probably extended to include other
collective degrees of freedom and symmetry restorations, are able to
produce flat energy differences instead of the parabolic behavior found in
Figs.~\ref{Masses_AW}-\ref{Masses_AW_ZZ} (and in
Refs.~\cite{PRC_73_034322_2006,PLB_668_420_2008,PRL_102_242501_2009,PRC_81_014303_2010}). These
tasks are highly demanding, both the refit of the interaction and the
inclusion of, for example, pairing
fluctuations~\cite{PRC_88_064311_2013},
triaxial~\cite{PRC_78_024309_2008,PRC_81_064323_2010,PRC_81_044311_2010,PRC_90_034306_2014}
and/or octupole~\cite{RMP_68_349_1996,PRL_80_4398_1998} shapes with
the corresponding symmetry restorations and configuration mixing. On
the other hand, it is possible that the central, spin-orbit and
density-dependent terms of the starting Gogny interaction have to be
modified including, for instance, explicit tensor
terms~\cite{PRL_97_162501_2006,PRC_86_054302_2012}.

Finally, since the present calculations are not fully converged, neither other degrees of freedom are taken into account -triaxiality, octupolarity, etc.- RMS values for D1S and/or D1M parametrizations given in Table~\ref{Tab1} should be considered as a qualitative description of the effect on the masses produced by the different approaches rather than the final values.

\subsubsection{Two particle separation energies and shell gaps}

Most of the times, the relevant quantities for calculating reaction
rates, $Q$-values, etc., with astrophysical interest are not the
absolute energies shown in the previous section but energy differences
between those masses. We analyze now its performance on two-nucleon
separation energies ($S_{2n},S_{2p}$), since the present GCM method
with the Gogny functional is not well developed for computing
odd-nuclei yet:

\begin{eqnarray}
S_{2n}(Z,N)&=&E(Z,N-2)-E(Z,N)\nonumber\\ 
S_{2p}(Z,N)&=&E(Z-2,N)-E(Z,N)
\label{s2n}
\end{eqnarray}  

These quantities are plotted in Figs.~\ref{S2n_AUDI}-~\ref{S2p_AUDI}
and analyzed quantitatively in Table~\ref{Tab1}. The overall behavior
of the experimental values is quite well reproduced. The RMS
deviations from experimental values for $S_{2n}$ and $S_{2p}$ are much
better than in the masses for both parametrizations. However, we
observe important local differences between the experimental data and
theory. On the one hand, the experimental curves are much smoother,
having always for a constant number of protons (neutrons) a continuous
decrease in the $S_{2n}$ ($S_{2p}$) when adding neutrons
(protons). This is not the case for the theoretical data, where jumps
and crossings between isotonic (isotopic) chains in the $S_{2n}$
($S_{2p}$) separation energies are shown. This drawback is obtained
all over the nuclear chart independently on the parametrization and is
a bit larger in the GCM approach than in the HFB result. Its origin could be related to a convergence problem discussed in Sec.~\ref{Theo} and/or the lack of other degrees of freedom such as triaxiality or octupolarity.

Finally, we analyze the neutron and proton shell gaps associated to magic numbers by zooming in the $S_{2n}$ energies 
for $N=20,\,22,\,28,\,30,\,50,\,52,\,82,\,84,\,126,\,128$ isotonic chains and the $S_{2p}$ energies 
for $Z=20,\,22,\,28,\,30,\,50,\,52,\,82,\,84$ isotopic chains in Figs.~\ref{shell_gaps_neutrons}-~\ref{shell_gaps_protons}.
We observe first that the different parametrizations (D1S and D1M) provide for this set of nuclei very similar results. Furthermore, the mean field approach tend to predict larger shell gaps than the experimental ones  except for $N,Z=28$ magic numbers, where the agreement with the experimental data is very good.
Correlations beyond the static mean field tend to reduce these gaps, almost matching the experimental results in $N,Z=20$ and $Z=50$. However, the reduction is not enough to
reproduce the actual gaps in $N=50$, 82, 126 nor $Z=82$. Therefore, the shell
quenching obtained by including BMF effects reported by Bender et
al. in Ref.~\cite{PRC_78_054312_2008} and by Delaroche et al. in
Ref.~\cite{PRC_81_014303_2010} is only partially reproduced here. In
addition, the latter results are much smoother than the results shown
in Figs.~\ref{S2n_AUDI}-\ref{S2p_AUDI}. As it was mentioned above (see Fig.~\ref{D1S_5DCH}), the amount of BMF correlation energies obtained by the present SCCM method is larger than the one obtained by solving the 5DCH model, even though in the latter the triaxiality is included. Furthermore, calculations reported in Refs.~\cite{PRC_73_034322_2006,PRC_78_054312_2008} are very similar to the ones presented here except for the use of GOA (and topGOA) approximations. Therefore, the smoothness of the $S_{2n}$
($S_{2p}$) separation energies in such calculations could be related to the GOA
approximations used in those references. Nevertheless, a better convergence and the addition of other degrees of freedom are needed to check whether the present interactions can reproduce the smoothness of the experimental data.

\subsubsection{$2_{1}^{+}$ excitation energies}

We finally compare the $2^{+}_{1}$ excitation energies obtained with
our present GCM global calculations with the experimental data
compiled in Ref.~\cite{ADNDT_78_1_2001}. It is important to note again that
the convergence of the results are not fully guaranteed,
especially in heavy nuclei where a working basis including eleven
major oscillator shells could be too small (see Sec.~\ref{Theo} and Fig.\ref{Fig_convergence}, bottom panel). Therefore, the values
obtained with this reduced space must be taken with caution. In any
case, we consider these results relevant to extract a global
performance of the method, and, in particular, to compare the results
provided by the two Gogny parametrizations for this observable. We can
also compare with the results of similar studies already reported with
the Skyrme SLy4~\cite{PRC_75_044305_2007} (with a topological gaussian
overlap approximation in the angular momentum projection and a limited
number of intrinsic states in the GCM) and Gogny
D1S~\cite{PRC_81_014303_2010,PRL_99_032502_2007} within the 5DCH framework. 
However, in the latter cases an educated selection of nuclei where the method is better suited was made and sets of 359~\cite{PRC_75_044305_2007}, 519~\cite{PRL_99_032502_2007} and 513~\cite{PRC_81_014303_2010} nuclei were chosen in these papers.

Both the experimental values and the results of the GCM calculations
for the $2^{+}_{1}$ energies are shown in Fig.~\ref{e2_d1s_d1m}. We
observe clearly an enhancement of the excitation energies at the
proton and neutron magic numbers. Additionally, the $2^{+}_{1}$
excitation energies corresponding to $N=20$, 28, 50, 82 and 126
isotonic chains and Ca, Ni, Sn and Pb isotopes are clearly above the
rest both experimentally and in the calculations. On the other hand,
we see that the two parametrizations provide almost identical results
for this observable even though they behave very different for
masses. The global behavior of the experimental data is well
reproduced although the calculations show a less smooth behavior and a
systematic overestimation of the experimental values. 

\begin{figure}[htb]
\begin{center}
\includegraphics[width=1.0\columnwidth]{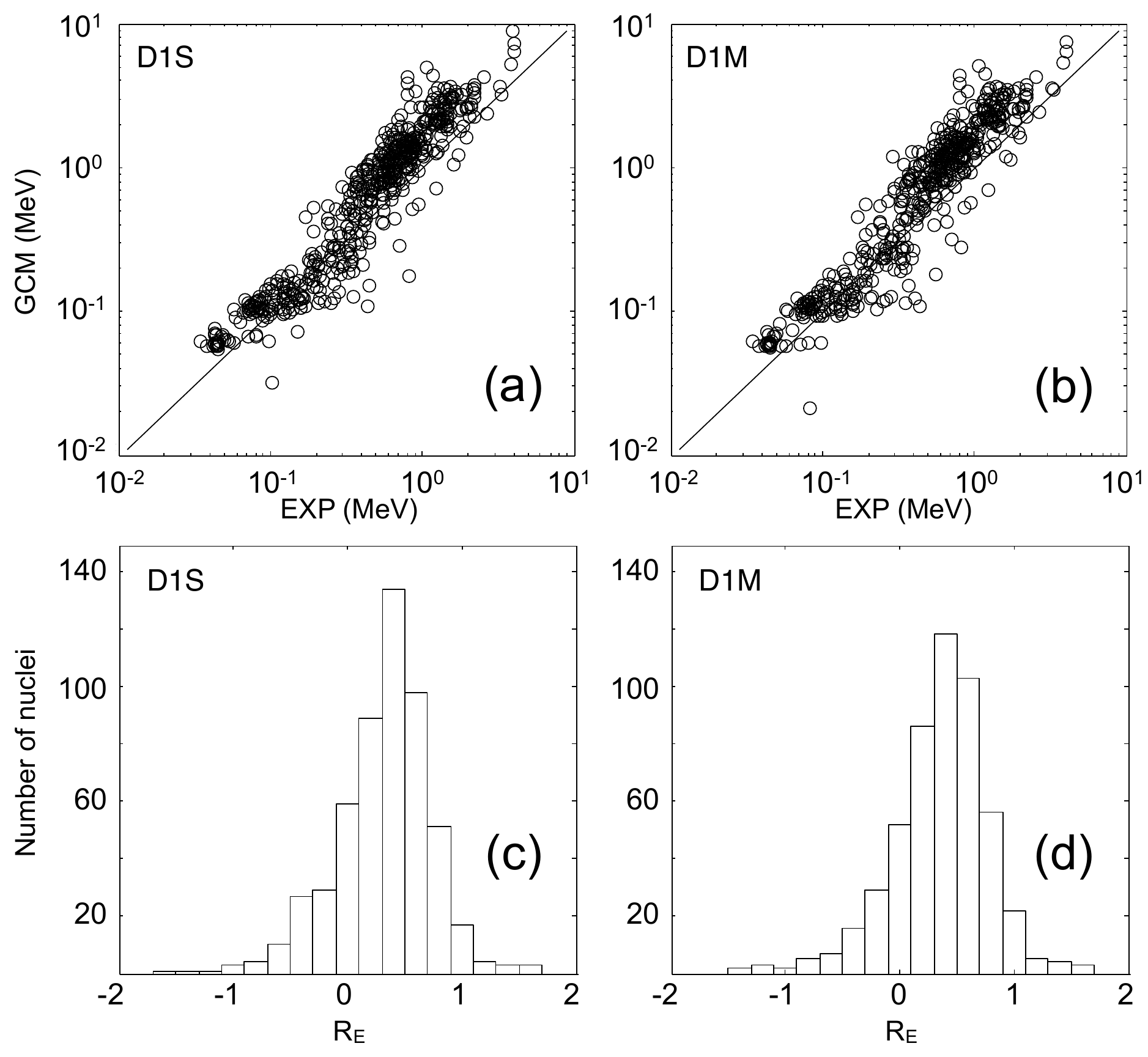}
\caption{GCM versus experimental $2^{+}_{1}$ excitation energies for
  (a) D1S and (b) D1M parametrizations. (c)-(d) Histograms of the logarithmic
  errors of the $2^{+}_{1}$ states.\label{e2_exp}}
\end{center}
\end{figure}

The latter drawback can be better seen in Fig.~\ref{e2_exp}(a)-(b),
where the theoretical values versus the experimental energies are
represented for D1S and D1M parametrizations respectively. Although we
observe a clear correlation between the two quantities, the
theoretical predictions tend to overestimate the empirical values. We
do not find significant differences between the D1S and D1M
parametrizations. To better estimate quantitatively the differences
with the experimental results we follow the analysis performed in
Refs.~\cite{PRC_75_044305_2007,PRL_99_032502_2007,PRC_81_014303_2010}. In these works, the
so-called logarithmic error -$R_{E}$- and its standard deviation
-$\sigma_{E}$- from the average -$\bar{R}_{E}$- are defined as: 

\begin{eqnarray}
R_{E}&=&\log\left[E(2^{+}_{1})_{\mathrm{th}}/E(2^{+}_{1})_{\mathrm{exp}}\right]\\
\sigma_{E}&=&\langle\left(R_{E}-\bar{R}_{E}\right)^{2}\rangle^{1/2}
\end{eqnarray}  

In Fig.~\ref{e2_exp}(c)-(d) we show histograms representing the number
of nuclei with a given value of $R_{E}$ for D1S and D1M
parametrizations. We find rather symmetric distributions with mean
values and standard deviations of $\bar{R}_{E}=0.32 (0.35)$ and
$\sigma_{E}=0.42 (0.43)$ respectively for D1S (D1M). Similar results
are obtained with Skyrme SLy4~\cite{PRC_75_044305_2007} with a more restricted set of nuclei (see
Table~\ref{Tab2}). This systematic drift towards larger values is a
consequence of the variational method. In this framework, the ground
state $0^{+}_{1}$ are favored with respect to other states, pulling
down its energy and stretching the final spectra. The inclusion of
additional degrees of freedom such as triaxiality, time-reversal
symmetry breaking and quasiparticle excitations, which are more
relevant in the excited states than in the ground state, allows a
variational description of these excited states and the excitation
energies can be reduced. For example, local studies with a GCM method
including triaxial angular momentum
projection~\cite{PRC_81_064323_2010,PRC_78_024309_2008,PRC_81_044311_2010,PRC_90_034306_2014}
or pairing fluctuations~\cite{PRC_88_064311_2013} have already shown
this effect.  On the other hand, quantum number projections and GCM
can modify significantly both the equilibrium deformation of the
system ($^{32}$Mg is the paradigm for this
effect~\cite{NPA_709_201_2002}) and the collective masses. Since these
parametrizations are not fitted with using the BMF many-body states of
this work, the MF and BMF collective behavior can be different. For
instance, angular momentum projection could overestimate a quadrupole
deformation that could be correct at the HFB level. Therefore, the
excitation energies should be taken into account in a future refit of
the interaction. 

We finally comment on the smaller deviation and dispersion obtained
with Gogny D1S using the 5DCH method~\cite{PRC_81_014303_2010} shown
in Table~\ref{Tab2}. In this case, the inclusion of the triaxial
degree of freedom and the fit of the inertia parameters with a
cranking approximation can improve the description of the $2^{+}_{1}$
excitation energies. However, since 5DCH does not include either
particle number or angular momentum projection, spurious contributions
of MF states with the wrong quantum numbers can produce a collapse of
the excited states~\cite{PRC_88_064311_2013}. In any case, further
analyses, beyond the scope of this work, are needed to compare the
5DCH approach with a full SCCM approximation. 

\section{Summary and outlook}
\label{Summ}

In this work, correlation energies, total energies, two neutron
separation energies and $2^{+}_{1}$ excitation energies for a large
set of even-even nuclei along the nuclear chart have been presented. The
theoretical results have been obtained by using self-consistent mean
field (HFB) and beyond-mean-field (BMF) methods, including 
particle number and angular momentum restorations and axial
quadrupole shape mixing without gaussian overlap approximations. The underlying interaction used in all of the
calculations was Gogny with two different parametrizations, D1S and
D1M. The convergence of the results as a function of the number of harmonic oscillator shells included in the basis has been analyzed. Hence, the convergence of the total energies are not fully guaranteed and extrapolation schemes to infinite bases should be implemented in the near future. Nevertheless, a large number of harmonic oscillator shells, $N_{s.h.o.}=19$,  has been used for the HFB part, showing a better performance than the usual rule of having a number of single particle states equal to eight times the larger number among the protons and neutrons of a given nucleus. Additionally, the calculated BMF correlations and particle separation energies are less dependent on the number of harmonic oscillator shells than the total energies.
Concerning those BMF correlations, both parametrizations, D1S and D1M, give similar correlation energies with
respect to their mean field solutions.
 
Compared to the experimental data, the D1S parametrization shows a
symmetry energy problem which produces a lack of binding energy in
neutron rich systems. This fact is reflected in a drift in the
difference between theoretical and experimental energies for nuclei
with a neutron excess which is corrected in the D1M
parametrization. However, strong shell effects (stronger as a function
of the neutron number) are still present in both realizations and
parabolic instead of flat energy differences are found between two consecutive
magic numbers. BFM correlations tend to reduce these differences since
rotational and vibrational corrections are larger in the mid-shell
than in closed shell nuclei, but this reduction is not sufficient to
remove the difference between the experimental and theoretical values.

Energy differences such as two-neutron separation energies are in a
better agreement with experimental data but still not
satisfactory. Hence, BMF correlations bring the theoretical predictions towards the experimental values for $N,Z=20$ shell gaps. However, some problems like artificial jumps, crossing and
overestimation of the shell gaps for heavier are found in the calculations and
they are not fully corrected by including BMF effects.

Additionally, we have reported the results for $2^{+}_{1}$ excitation
energies calculated with $N_{s.h.o.}=11$ shells. We have obtained
similar results for D1S and D1M parametrizations, having a systematic
stretching of the $2^{+}_{1}$ energies which is more related to the
method used to solve the many-body problem rather than the
parametrization itself.

As an outlook, some improvements should be taken into account in the
near future, in particular: 
\begin{enumerate}
\item Convergence of the results with the properties of the harmonic
  oscillator working basis. 
\item Triaxial and other degrees of freedom should be included
  explicitly in the calculations. 
\end{enumerate}
In the long-term, a new parametrization of the Gogny interaction (or any other type of energy density functional) including in the fitting procedure fully converged SCCM corrections with triaxial and octupolar states is desirable.
 
 Finally, new experimental data would be very helpful to constrain
further the current and future models.

\section*{Acknowledgements}
We acknowledge the support from GSI-Darmstadt and CSC-Loewe-Frankfurt computing facilities.
This work was supported by the BMBF-Verbundforschungsprojekt number
06DA7047I, the Ministerio de Econom\'ia y Competitividad-Programa
Ram\'on y Cajal 2012 number 11420 and the Helmholtz Association
through the Nuclear Astrophysics Virtual Institute (VH-VI-417).


\begin{thebibliography}{10}
\bibitem{footnote}
Although the working basis is spherical, the HFB states, defined by the transformation given in Eq.~\ref{HFBtrans}, can be deformed.
\end{thebibliography}
\end{document}